\documentclass[a4paper,fleqn,usenatbib]{mnras}
\newcommand{\ul}{ULIRG}
\newcommand{\li}{LIRG}
\newcommand{\sfg}{star-forming galaxy}
\newcommand{\sfgs}{star-forming galaxies}
\newcommand{\lirul}{$L_{\rm{IR}}\geq10^{12} L_{\odot}$}
\newcommand{\lirli}{$10^{11} L_{\odot}\leq L_{\rm{IR}}<10^{12} L_{\odot}$}
\newcommand{\lirsf}{$L_{\rm{IR}}<10^{11} L_{\odot}$}
\newcommand{\smx}{$70\micron$/X-ray\ }

\usepackage{times}
\usepackage{epsfig}
\usepackage{amsmath}
\usepackage{natbib}
\usepackage{graphicx}
\usepackage{footmisc}
\usepackage[usenames,dvipsnames]{xcolor}

\title{Role of AGNs in the Luminous Infrared Galaxy phase since $\lowercase{z} \sim 3$}
\author[M-Y Lin, Y. Hashimoto, and S. Foucaud]{Ming-Yi Lin$^{1,2}$\thanks{E-mail:
699440083@ntnu.edu.tw, mingyi@mpe.mpg.de}, Yasuhiro Hashimoto$^2$ and S\'ebastien Foucaud$^{3,4,2}$\\
            $^1$Max-Planck-Institut fŸr Extraterrestrische Physik (MPE), Giessenbachstrasse 1,85748 Garching, Germany \\
            $^2$Department of Earth Sciences, National Taiwan Normal University, No.88, Tingzhou Road, Sec. 4, Taipei 11677, Taiwan \\
            $^3$Center for Astronomy \& Astrophysics, Department of Physics \& Astronomy, Shanghai JiaoTong University, 800 Dongchuan Road, Shanghai 200240, China \\
            $^4$Institute of Astronomy and Astrophysics, Academia Sinica, P.O. Box 23-14, Taipei 10617, Taiwan  \\}

\begin{document}           
            
\date{Accepted 2015 November 20.  Received 2015 November 19; in original form 2013 January 12}

\pagerange{\pageref{firstpage}--\pageref{lastpage}} \pubyear{2002}

\label{firstpage}

\maketitle

\begin{abstract}
In order to understand the interactions between active galactic nuclei (AGN) and star formation during the evolution of galaxies, we investigate 142 galaxies detected in both X-ray and 70\micron\ observations in the COSMOS (Cosmic Evolution Survey) field. All of our data are obtained from the archive, X-ray point source catalogs from Chandra and XMM-Newton observations; far-infrared 70\micron\ point source catalog from Spitzer-MIPS observations. 
Although the IRAC [3.6\micron]-[4.5\micron] vs. [5.8\micron]-[8.0\micron] colours of our sample indicate that only $\sim$63\% of our sources would be classified as AGN, the ratio of the rest-frame 2-10 keV luminosity to the total infrared luminosity (8-1000\micron) shows that all of the sample has comparatively higher X-ray luminosity than that expected from pure star-forming galaxies, suggesting the presence of an AGN in all of our sources.
From the analysis of the X-ray hardness ratio, we find that sources with both 70$\micron$ and X-ray detection tend to have a higher hardness ratio relative to the whole X-ray selected source population, suggesting the presence of more X-ray absorption in the 70$\micron$ detected sources.
In addition, we find that the observed far-infrared colours of 70$\micron$ detected sources with and without X-ray emission are similar, suggesting the far-infrared emission could be mainly powered by star formation.

\end{abstract}


\section{INTRODUCTION}

Observational and theoretical evidences over the past decades strongly suggest the existence of supermassive black hole (SMBH) within the spherical bulge of galaxies. The tight connection between the SMBH mass and bulge properties, such as bulge luminosity or stellar velocity dispersion, indicates a scenario of co-evolution of central black hole and its host galaxy \citep{Kormendy1995, Magorrian1998}. Galaxy merger is a plausible interpretation to explain the concomitant growth of black hole and host galaxy under the $\Lambda$CDM model \citep{Treister2010}. 
As tidal effects distort the morphology of galaxies during mergers, episodes of extreme star formation can occur; therefore interstellar medium (ISM) is abundantly produced, either from supernova explosions or stellar wind of evolved stars. Accompanying this morphological effect, merger not only reduces angular momentum but also induces gas accumulation around circumnuclear region \citep{BH1991}. The large amount of dust and molecular gas around the central region funnels onto the SMBH in the galactic nucleus and consequently triggers a phase of active galactic nuclei (AGNs), with its powerful electromagnetic radiation spanning from radio to gamma-ray. In the later galaxy evolution stages, the merger scenario implies that the radiation-pressure or kinematic-wind feedback from AGN may quench the star formation, expelling the remnant gas and dust. At the same time, central AGN turns into an optically unobscured quasar, a class of luminous AGN. 
Final outcome of the galaxy evolution is a red elliptical galaxy \citep[and references therein]{Hopkins2008}.

\citet{Sanders1988} first proposed a connection between the Ultra Luminous Infrared Galaxies (\ul s; \lirul) and quasars as two successive snapshots of merger event between two gas-rich spiral galaxies. Indeed, the analysis of the IRAS 1 Jy \ul s survey revealed a large fraction of interacting galaxies at $z<0.1$ in favor of such merger-\ul\ scenario \citep{Borne2000, Cui2001,Veilleux2002}. Besides, the merger fraction has been shown to increase as infrared luminosity increases, merging galaxies being the prevalent population amongst \ul s to redshift z $\sim$ 1 \citep{Shi2009}. 
However at higher redshift z $\sim$ 2, the merging galaxy fraction drops slightly and non-interacting disk galaxy fraction increases \citep{Kartaltepe2012}.
On the other hand, the merger-AGN connection is more difficult to observe. Although, in the local universe, optical luminous quasars are considered to be related to post-starburst mergers \citep{Canalizo2001,Bennert2008}; at 1.5 $<$ z $<$ 2.5, AGNs are more likely to occur inside of normal disk or bulge galaxies instead of interacting systems, suggesting that secular processes (such as bars or nucleus rings) may trigger nuclear activity instead of major-merger \citep{Schawinski2011,Kocevsick2012}. 
The connection between ULIRGs and AGNs has been investigated for the past few years. 
The AGN fraction seems to rise with increasing infrared luminosity \citep{Genzel2000,Kartaltepe2010} as the AGN activity likely influences the host galaxy infrared spectral energy distribution (SED). Galaxies with higher mid-infrared flux ratio (i.e. F25/F60 $\geq$ 0.2) may be classified as ÒwarmÓ ULIRGs, which are supposed to be dominated by AGN. Galaxies with lower mid-infrared flux ratio (i.e. F25/F60 $\leq$ 0.2) may be classified as ÒcoldÓ ULIRGs, which are supposed to be dominated by star formation \citep{Sanders1988-2,Veilleux2009}.
 
In order to determine the exact role of the AGN in the overall galaxy evolution, it is essential to separate the AGN contribution from the star formation activity. 
To decouple these two components, three distinct methods can be used:\\
(i) {\it Specific emission lines ratio} - Fine structure radiation in optical and infrared wavelength are reliable tracers of excitation state of the ISM in galaxies. A classification based on two independent line ratios can help to distinguish AGN from star forming galaxies. For instance, it is possible to separate the main mechanism responsible for the excitation of the gas, between normal HII regions, objects photoionized by AGN, and shockwave heating by star formation, using [OIII]5007\AA/H$\beta$ versus [NII]6584\AA/H$\alpha$ diagram \citep[][BPT]{BPT}.
Unfortunately, galaxy mergers suffer from dust obscuration, especially at optical wavelengths, reducing the efficiency of such diagnostics; although a complete misclassification of AGN is unlikely \citep{Veilleux20021, LaMassa2012}.

With the advent of the infrared spectrometer on the Infrared Space Observatory (ISO) and the Spitzer telescope, other emission lines were identified for AGN diagnostics in the mid-infrared with less extinction. For example, the line ratio [O IV]25.9\micron/[Nell]12.8\micron\ or [NeV]14.3\micron/[Nell]12.8\micron\ is a reliable indicator to separate AGNs from \sfgs\ \citep{Genzel1998, Armus2006}.
The different Polycyclic Aromatic Hydrocarbon (PAH) emission features at 6.2\micron\, 7.7\micron\, and 11.3 \micron\ are other reliable star formation tracers, with weaker PAH emission corresponding to higher AGN contribution \citep{Voit1992,Veilleux2009};\\
(ii) {\it Continuum slope} - The cold dust emission in the far-infrared band is mainly due to star formation. In contrast, hot dust components around 10$\micron$ are considered to be associated with AGN activity. As the influence of the AGN becomes significant in the galaxy, the infrared spectral indexes start to be dominated by a power-law continuum $\sim$ $10\micron$ \citep{AH2006,Donley2007}. Therefore, the mid-infrared to far-infrared ratio is a simple measure to quantify the relative contributions of AGN and star formation \citep{Veilleux2009};\\   
(iii) {\it Spectral energy distribution fitting} - With the development of deep multi-wavelength photometric surveys, panchromatic SED studies are becoming increasingly popular. Theoretical or empirical templates are commonly used to fit the SED and determine the photometric redshifts of galaxies. Multi-component fitting, including the contribution from stars, dust, and AGNs is now possible due to very broad photometric coverage. In the case of heavily obscured AGNs, although star formation of host galaxy dominates the flux at most wavelengths, a bump in mid-infrared produced by the circumnuclear dust radiation can be an indicator of an obscured AGN. Therefore, identifying hot dust component from the galaxy SED is a reliable way to determine the AGN contribution to total infrared luminosity \citep{Mullaney2011,Pozzi2012}.

Although the different methods may not agree completely on defining a pure sample (i.e. pure-AGN or pure starburst galaxy), they all provide evidences that, for the major part of the samples, AGN and star formation occur concomitantly. 
What is essential then is to understand the mutual influence on the properties of AGNs and their host \sfgs. We try to understand whether the dust and gas from star formation in host galaxy could obscure central AGN or the presence of AGN radiation could change the far-infrared SED.

The SEDs of Luminous Infrared Galaxies (\li s; \lirli) and \ul s peak in the range of 40-200\micron\ \citep{Sanders1996}. In fact, the 70\micron\ Spitzer band is ideal to unambiguously trace \sfgs\ in the local universe as it is little affected by PAH features, silicate feature, and stellar flux. Central AGNs are usually identified by their strong X-ray continuum emission. Indeed, high energy X-ray photons emitted by hot corona of accretion disk \citep[e.g.,][]{Haardt1993} around the central black hole are usually less absorbed by dust and gas from host galaxies compared to lower energy UV-photons. One noticeable exception is Compton thick objects ($N_{H} > 10^{24}$ cm$^{-2}$) for which even hard X-ray photons are absorbed. 
Compton-thick AGNs are not an uncommon occurrance among \ul\ population. The detection of Fe K$\alpha$ lines and large amount of neutral hydrogen absorption in hard X-ray spectrum support the presence of Compton-thick AGN inside \ul\ (e.g. Mrk 273 in \citet{ Balestra2005} paper).

Several studies have investigated the nature of the 70\micron\ galaxy population. \citet{Patel2011} led a spectroscopic follow-up in optical wavelength of 70\micron\ galaxies selected from the Spitzer Wide-area Infrared Extragalactic (SWIRE) Legacy Survey. Their results suggest that, for galaxies not dominated by QSO, most of the IR photons are emitted by star-forming regions, while contribution of emission by AGN dusty torus is negligible. Furthermore, \citet{Symeonidis2010} investigated the SEDs of 61 70$\micron$-selected galaxies from the 0.5 deg$^{2}$ wide Extended Groth Strip (EGS) field and concluded that, even for galaxies displaying powerful hard X-ray emission originating from AGNs, cold dust emission templates are still required to explain the observed strong far-infrared luminosity. To reveal a potential starburst-AGN connection, \citet{Trichas2009} used 28 X-ray sources with 70\micron\ detection and applied a statistical Kolmogorov-Smirnov test (K-S test) to assess any difference in hardness ratio between 70\micron\ detected X-ray galaxies and the whole X-ray population in the redshift interval 0.5 $<$ z $<$1.3. However, the K-S test shows a none significant probability, in contrast to the prediction of AGN/starburst co-evolution models \citep{Hopkins2008}.

In this paper, we will unveil the role of AGN in the infrared (ultra)luminous phase of galaxy evolution by comparing the properties of an X-ray detected subsample and its parent 70$\micron$-selected sample. Our main 70\micron\ galaxy catalog was published by \citet{Kartaltepe2010}, which includes total infrared luminosity, multi-wavelength photometry and redshift. We describe the data used for this work in section 2. As we are interested in the physical properties of AGN and their far-infrared selected host galaxy, we have to cross-match X-ray and 70\micron\ datasets. In section 3, we depict our matching method as well as the methods used to estimate and calibrate different physical parameter measurements. We present our results in section 4, and engage in detailed discussions in section 5. Throughout this paper, we adopt a standard cosmological model with the following parameters: $H_{0}$ = 70 km s$^{-1}$ Mpc$^{-1}$, and $\Omega_{M}$ = 0.3, $\Omega_{\Lambda}$ = 0.7. Unless otherwise stated, all magnitudes in this paper are in the Vega system.

\section{DATA}

The Cosmic Evolution Survey (COSMOS) is the largest Treasury program using Hubble Space Telescope imaging a $\sim$2 deg$^{2}$-wide equatorial field in the optical F814W filter (approximately corresponds to I band) with the Advanced Camera for Surveys (ACS) instrument \citep{Scoville2007,Scoville2007-2,Koekemoer2007}. Several large follow-up campaigns using ground-based and space-based telescopes have produced a comprehensive photometric and spectroscopic dataset across a whole spectrum \citep{Capak2007,Hasinger2007,Sanders200724,Scott2008}\footnote{c.f. the COSMOS Special Issue of the Astrophysical Journal Supplement Series, in September, 2007.}. 
\citet{Kartaltepe2010} published a reliable 70$\micron$ selected galaxy sample (1503 sources) with signal-to-noise (S/N) $>$ 3 from Spitzer data in the COSMOS field and provided their corresponding redshift, total infrared luminosity, and multi-wavelength photometry. In this study, we attempt to find the X-ray counterparts of 70$\micron$ selected galaxies. We use both XMM-Newton and Chandra observations in the COSMOS field, taking advantage of the wider observational field of the XMM-Newton survey and the deeper observational depth of the Chandra survey to maximize the total number of X-ray point source counterparts in the 70$\micron$ selected galaxy sample.
In the following subsections, we will give a brief description of infrared and X-ray datasets.

\subsection{Spitzer-COSMOS}
\label{sec:scosmos}

Spitzer-COSMOS (S-COSMOS) is a Legacy program designed to cover the COSMOS field with deep Spitzer observations in IRAC four bands (i.e. 3.6, 4.5, 5.8, and 8.0\micron) and MIPS three bands (i.e. 24, 70, and 160\micron). The initial Cycle 2 program conducted in 2006 consisted of a deep IRAC survey and a narrow-field MIPS survey. The total integration time of the deep IRAC survey is 166 hours, reaching $5\sigma$ sensitivities of 0.9, 1.7, 11.3, and 14.6 $\mu$Jy, in 3.6, 4.5, 5.8, and 8.0\micron\ band, respectively \citep{Sanders200724}. A complementary MIPS deep observation of 450 hours has been conducted in Cycle 3 ensuring accurate 70 and 160\micron\ flux density measurements. The median exposure times were $\sim$ 3400s, 1350s, and 270s for 24$\micron$, 70$\micron$, and 160\micron\ bands, respectively, corresponding to $5\sigma$ depths of $\sim$ 0.08, 8.5, and 65 mJy \citep{LeF2009,Frayer2009,Kartaltepe2010}.

\subsection{X-ray dataset: XMM-Newton \& Chandra}
\label{sec:xray}

The COSMOS field has been covered with observations from both XMM-Newton (XMM) and Chandra observatories. 

A 2 deg$^{2}$ contiguous survey has been conducted with XMM to reach medium depth ($\sim$60 ks).
COSMOS X-ray catalog made by XMM observation provides three different energy bands. Based on the logN-logS relationship, the flux limit of 0.5-2 keV (soft band), 2-10 keV (hard band), and 5-10 keV (ultra-hard band) dropped to 7.2 x 10$^{-16}$ ergs cm$^{-2}$ s$^{-1}$, 4.0 x 10$^{-15}$ ergs cm$^{-2}$ s$^{-1}$, and 9.7 x 10$^{-15}$ ergs cm$^{-2}$ s$^{-1}$, respectively \citep{Hasinger2007,C2009XMM}. In total, 1887 point-like sources were identified in the XMM-COSMOS survey. \citet{Brusa2010} has released a cross matched catalog between these sources and optical counterparts (I band, catalog of \cite{Capak2007}), providing redshifts (spectroscopic and photometric), UV to 24\micron\ photometry, and hardness ratio.

The Chandra-COSMOS (C-COSMOS) Survey images a $\sim$0.9 deg$^{2}$ field in the central part of the original COSMOS field, with the effective exposure time $\sim$160 ks in center 0.5 deg$^{2}$ and $\sim$80 ks in outer 0.4 deg$^{2}$. 
C-COSMOS catalog provides three different energy bands, 0.5-2 keV (soft band), 2-7 keV (hard band), and 0.5-7 keV (full band), with corresponding flux limits of 1.9 x 10$^{-16}$ ergs cm$^{-2}$ s$^{-1}$, 7.3 x 10$^{-16}$ ergs cm$^{-2}$ s$^{-1}$, and 5.7 x 10$^{-16}$ ergs cm$^{-2}$ s$^{-1}$, respectively. In total, 1761 point-like sources have been extracted in the in C-COSMOS field \citep{elvis2009}.

\section{METHOD}

\subsection{Matched X-ray and far-infrared point source catalog}
\label{sec:smx}

\begin{table*}
\begin{center}
  \caption{Samples properties and matching results}
  \vspace{-0.3cm}
    \begin{tabular}{cp{0.3cm}lp{0.15cm}rl*{3}{c}} \hline               
  Catalog      & \multicolumn{2}{c}{Area $^{(a)}$}    & \multicolumn{3}{c}{Flux limits $^{(b)}$}  & $N_S$ $^{(c)}$ & $N_{70\micron}$ $^{(d)}$  & References  $^{(e)}$  \\ 
        &  \multicolumn{2}{c}{in deg$^2$} & \multicolumn{3}{c}{in mJy  $^{(\dagger)}$} &  & & \\ 
        &  & & \multicolumn{3}{c}{or 10$^{-16}$ergs cm$^{-2}$ s$^{-1}$ $^{(\ddagger)}$} & &  & \\ 
  \hline 
  \hline
    Spitzer-COSMOS (S-COSMOS)  & \multicolumn{2}{c}{2.47} & & 8.5  $^{(\dagger)}$ & ($5\sigma$) & 1503    &  1503 & \citet{Frayer2009}  \\
    $70\micron$ galaxies catalog & & & & & & & & \& \citet{Kartaltepe2010} \\\hline
  XMM-COSMOS     & \multicolumn{2}{c}{2.13} & & 7.2  $^{(\ddagger)}$ & (Soft band) & 1797  & 108 &\citet{C2009XMM}   \\
 Point source catalog   & & & & 40 $^{(\ddagger)}$ & (Hard band) & & & \& \citet{Brusa2010}\\
        & & & & 97 $^{(\ddagger)}$ & (Ultra hard band) & & & \\\hline
  Chandra-COSMOS (C-COSMOS)  & \multicolumn{2}{c}{0.9$^{(f)}$}    & & 1.9 $^{(\ddagger)}$ & (Soft band) & 1761   & 92 & \citet{elvis2009} \\
  Point source catalog    &  &  & & 7.3 $^{(\ddagger)}$ & (Hard band) & &   & \\
        & & & & 5.7 $^{(\ddagger)}$ & (Full band) & & & \\\hline 
  XMM- \&  Chandra-COSMOS  & \multicolumn{2}{c}{0.9$^{(h)}$} & \multicolumn{3}{c}{$-$} & 2821 &142 & This work \\
 concatenated catalog $^{(g)}$  &  &  &  &  &  & & \\\hline 
        \end{tabular}
 \\
$^{(a)}$Effective area covered by the survey; $^{(b)}$ Flux limit of the catalog; $^{(c)}$Total number of sources; $^{(d)}$Number of sources cross-matched with the $70\micron$ catalog;  $^{(e)}$References describing the catalog; $^{(f)}$ central 0.5 deeper; $^{(g)}$Catalog concatenated from XMM- and Chandra-catalogs; sources matched between the two catalogs are not duplicated; the cross-match between this catalog and the $70\micron$ catalog constitute our primary "\smx" catalog; $^{(h)}$Note that the sky coverage of C-COSMOS is included in the XMM-COSMOS sky coverage.
\label{tab:smx}   
\end{center}
\end{table*}

The spatial resolution of the Spitzer Telescope at 70\micron\ is around 18$\arcsec$, much larger than the spatial resolution of XMM or Chandra X-ray observatory. Therefore, it is crucial to accurately detect the position of the 70\micron\ galaxies. \citet{Kartaltepe2010} have identified 1503 optical or near-IR counterpart to the 70\micron\ selected galaxies. We take advantage of the Altas resource from IRSA/COSMOS archive\footnote{NASA/IPAC infrared service archive of COSMOS project. http://irsa.ipac.caltech.edu/data/COSMOS} to match the optically or near-IR identified counterparts of 70$\micron$ selected galaxies with the XMM and Chandra point source catalogs, separately. 
We independently match 108 and 92 of our 70$\micron$-selected galaxies with counterparts from the XMM and the Chandra catalogs, using searching radii of 3" and 1", respectively. The searching radii are defined by the 90\% completeness of X-ray-to-Optical position offset distribution (c.f. upper panel in Figure 3 of \citet{Brusa2007} for XMM-COSMOS sample; Figure 6 of \citet{elvis2009} for C-COSMOS).
The median angular distances between the position of the X-ray source and the position of optical/near-IR identified counterpart of our $70\micron$-selected galaxies are $0.9\arcsec$ and $0.3\arcsec$ for the XMM and Chandra point source catalogs, respectively. 58 sources are identified in both X-ray catalogs, and a total number of 142 galaxies are assigned to both 70$\micron$ and X-ray measurements in the COSMOS field. We will refer to this sample as "\smx" galaxy catalog in the remaining part of this article. From the same datasets, \citet{Kartaltepe2010} had identified 154 X-ray detected $70\micron$ galaxies (refer to Figure 7 in their paper). The 12 extra objects compared to our own matched catalog resulted from a slightly less strict matching criterion used by \citet{Kartaltepe2010}. We speculate that these additional sources are more distant from the 70\micron\ source position. The number of sources from our different catalogs and matched catalogs are summarized in Table~\ref{tab:smx}.

\subsection{Redshift determination and total infrared luminosity}
\label{sec:zlir}

Several spectroscopic surveys were conducted to measure the redshifts of galaxies in the COSMOS field. The most extensive survey in this field is the zCOSMOS survey \citep{SLilly2007},  a deep spectroscopic survey conducted with the VLT-VIMOS multi-object spectrograph to  study the evolution of faint/obscured galaxies through direct analysis of their emission lines. Although the observations enable us to probe properties of faint objects, only the central 1~deg$^{2}$ is covered by zCOSMOS. Forty-nine of our \smx galaxies have their redshifts confirmed spectroscopically by zCOSMOS.
Observations using other facilities have been conducted on this field providing precise redshifts for fifty-three additional \smx galaxies \citep{sdss2009,trump2007,Kartaltepe2010,Prescott2006}. 
In total 102/142 (72\%) galaxies with X-ray and 70$\micron$ detection have spectroscopic redshift. Table~\ref{tab:reds} summarizes the origin and number of our targets with spectroscopic redshifts. 

For the remaining galaxies, we use the photometric redshift catalogs built for both non-AGN sources and XMM-selected sources \citep{llbert2009,Salvato2009} using the publicly available software Le PHARE\footnote{http://www.cfht.hawaii.edu/$\sim$arnouts/LEPHARE \label{fn:leph}} \citep{Arnouts1999,llbert2006}. The \smx galaxies that do not have spectroscopic redshift have been assigned a photometric redshift (40 out of 142 - 28\%).

We use the overlap between our spectroscopic and photometric redshift samples to investigate the quality of the later. The relation between photometric and spectroscopic redshift of \smx galaxies presents no bias and is comparable to overall 70$\micron$ galaxy with a scatter of $\sigma_{z} =$ $0.02 \times(1+z)$ \citep{Kartaltepe2010}. Bright sources have usually been assigned a spectroscopic redshift, with small error, while faint sources have been assigned a photometric redshift with larger error.
Incidentally, even though these two different methods for redshift measurement have different error sizes, the infrared and X-ray luminosity distribution inferred both by spectroscopically and photometrically determined redshfits have the same median value. Therefore, we conclude the difference in redshift measurements does not induce any systematic bias.

\begin{table}
  \caption{Confirmed spectroscopic redshift sample}
    \begin{tabular}{*{4}{c}} \hline               
     Telescope & Instrument &  $N$$^{(a)}$  & Reference      \\ 
  \hline \hline
  ESO-VLT  & VIMOS     & 49     & zCOSMOS  \\
    &      &      & \citet{SLilly2007} \\
  Magellan & IMACS & 26  & \citet{trump2007}   \\
  Sloan  & SDSS  & 15   & \citet{sdss2009}    \\
  Keck II & DEIMOS & 11   & \citet{Kartaltepe2010}    \\
  MMT & Hectospec   & 1  & \citet{Prescott2006}    \\\hline
    \end{tabular} 
    \\ 
 $^{(a)}$Number of \smx sources with spectroscpic redshift.
    \label{tab:reds}    
\end{table}
 
\citet{Kartaltepe2010} used Spitzer-MIPS (24, 70, 160$\micron$) data points to estimate the total infrared luminosity ($L_{\rm IR}$) for 70$\micron$ selected galaxies by fitting the SEDs with a $\chi^{2}$-minimization method using the Le PHARE\footref{fn:leph} SED fitting code and integrating the luminosity over the 8-1000$\micron$ wavelength range. Four infrared libraries have been used, including the SEDs of local star-forming galaxies and SEDs derived from radiative transfer models. The details are described in Sec. 4.2 of \citet{Kartaltepe2010}.
However, this catalog suffers from incompleteness in the 160$\micron$ band and lacks any direct photometric observation in the Rayleigh-Jeans part of the SED. 
There are 463 70$\micron$ sources with 160$\micron$ detections (463 out of 1503; 31{\%}), and most of them (417 out of 463; 89{\%}) have a single matched 160\micron\ source within the given MIPS error circle. 
For the 160\micron\ sources with multiple 70\micron\ counterparts, \citet{Kartaltepe2010} assigned the brightest 70\micron\ source as the counterpart of the 160\micron\ source.
\citet{Kartaltepe2010} assessed that measurements of $L_{\rm IR}$ without 160$\micron$ data are underestimated by 0.2 dex compared to measurements with 160$\micron$ data. Despite these caveats, we assume that this catalog is robust enough for our study, as the galaxies without 160$\micron$ photometry are equally spread between the samples of 70$\micron$ galaxies with or without X-ray detections.
Another caveat in this catalog is that the SED fitting does not take into account any AGN contribution at mid-infrared wavelengths. Therefore, for those 70$\micron$ galaxies having a significant contribution from an AGN hot dust component over 5-30$\micron$, $L_{\rm IR}$ may be contaminated by AGN emission. Since the SED of the hot dust itself drops dramatically beyond 30$\micron$ \citep{Lusso2013}, the AGN contribution to 8-1000$\micron$ emission is likely to be small. Other studies have investigated this issue, e.g., \citet{Pozzi2012} used luminous infrared galaxies at z$\sim$2 with Herschel observations in the Chandra Deep Field-South (CDF-S) field and concluded that the AGN contribution to 8-1000$\micron$ emission is only 5\%.

\begin{figure}
\begin{center}
  \includegraphics[width=84mm]{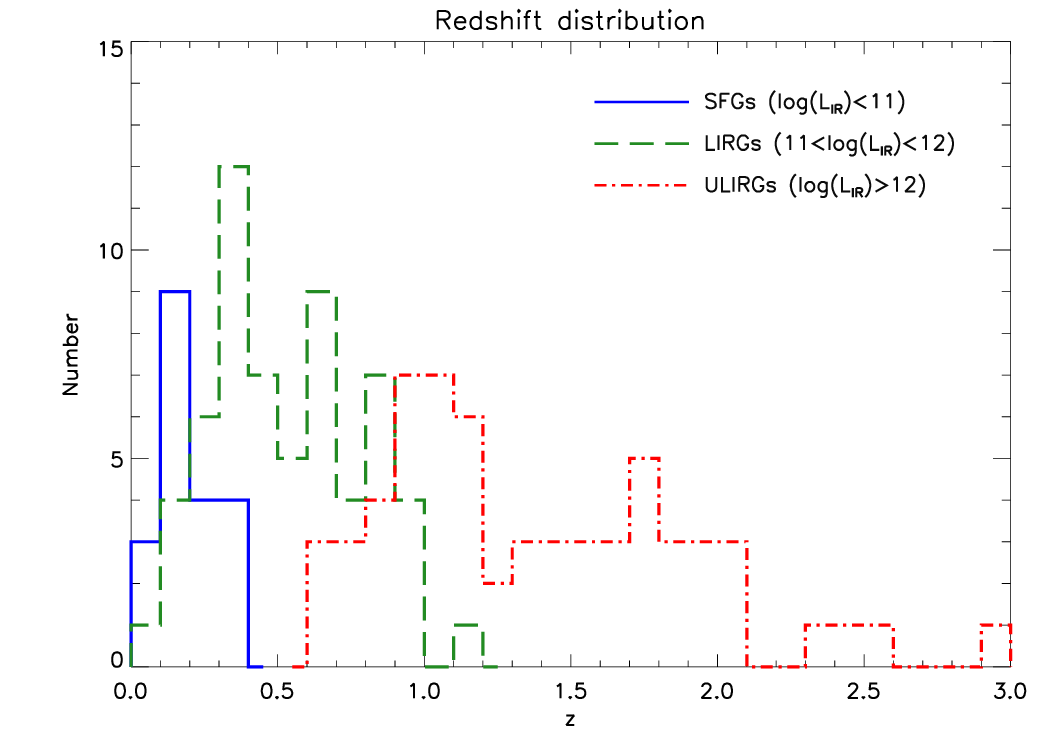}
    \caption{The redshift distribution of \smx galaxies. We divide our total sample into three distinct classes according to their total infrared luminosity: \sfgs\ (SFGs; \lirsf;  blue solid line) at a median redshift of $z \sim 0.168$, luminous infrared galaxies (\li s; \lirli;  green dash line) at $z \sim 0.518$ and ultra-luminous infrared galaxies (\ul; \lirul;  red dash-dot line) at $z \sim 1.268$.}
\label{fig:nz}
\end{center}
\end{figure}

We divide our \smx galaxy sample (142) into three distinct classes, according to their total infrared luminosity as \sfg\ (SFG, 20/142), luminous infrared galaxy (\li, 60/142), and ultra-luminous infrared galaxy (\ul, 62/142) with \lirsf, \lirli, and \lirul, respectively. Figure~\ref{fig:nz} shows the redshift distributions in the different infrared luminosity intervals: SFGs with median $z \sim 0.168$, \li s with median $z \sim 0.518$, and \ul s with median $z \sim 1.268$.

\subsection{Correction of X-ray properties}
\label{sec:hr}

\begin{figure}
      \includegraphics[width=84mm]{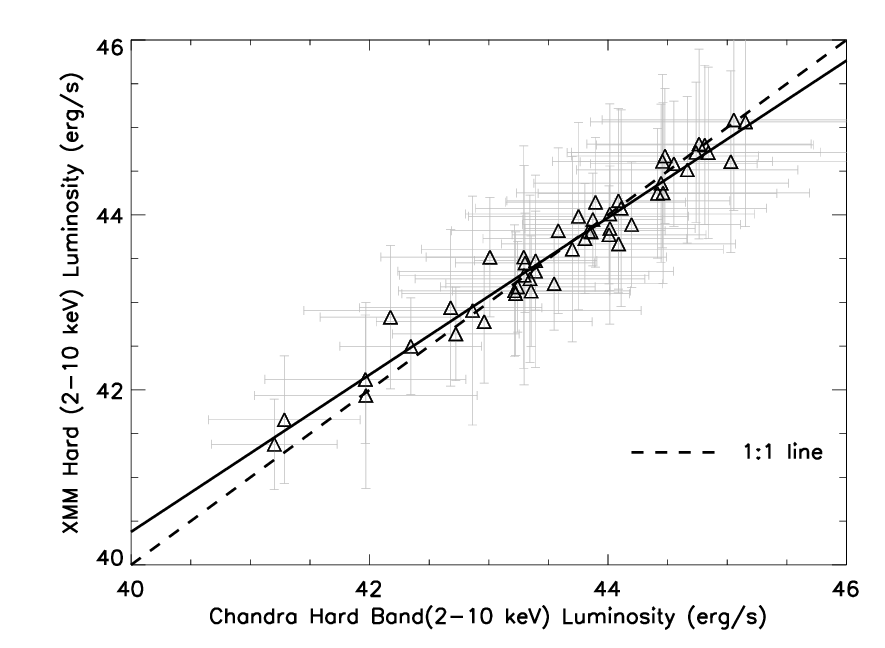}
         \caption{Hard X-ray luminosity from the XMM catalog \citep{Brusa2010} versus hard X-ray luminosity from the Chandra catalog \citep{elvis2009} for our \smx galaxies. A linear fit (see Equation~(\ref{eq:hleq})) is shown as the black solid line. The black dash line indicates 1:1 line.}
\label{fig:hlx}
\end{figure}

By combining both XMM and Chandra X-ray catalogs, we extend our total sample of \smx galaxies to 142 sources, 58 of which are listed in both X-ray catalogs. 
The Chandra and XMM catalogs use different energy bands to present hard X-ray count rate: 2-7 keV for Chandra and 2-10 keV for XMM. 
In both the XMM and Chandra catalogs, the hard X-ray count rate is converted (extrapolated) into the hard X-ray flux at 2-10 keV by assuming a power-law spectrum with spectral index $\Gamma$ and Galactic column density $N_{H}$. However, the two catalogs use different assumptions: XMM catalog used $\Gamma$ = 1.7 and Galactic column density $N_{H}$ = 2.5 $\times$ 10$^{20}$ cm$^{-2}$ \citep{C2009XMM}; Chandra catalog used $\Gamma$ = 1.4 and Galactic column density $N_{H}$ = 2.7 $\times$ 10$^{20}$ cm$^{-2}$ \citep{elvis2009}.
Because different assumptions have been applied to the X-ray catalogs, to make two catalogs consistent, we calculate the hard X-ray luminosity conversion equation by fitting a line to the hard X-ray luminosities for the 58 sources with detection in both catalogs. 
We denote that, in this section, the paired data are fitted to a linear equation (y = a$\times$x + b) by minimum $\chi^2$ method. The measured errors in observational data have been taken into account. The best-fitted linear model parameters [a,b] include the 1-$\sigma$ uncertainties. 
Figure~\ref{fig:hlx} shows the correlation of hard X-ray luminosity between XMM and Chandra.
The hard X-ray luminosity conversion equation is: 
\begin{equation}
log(L_{X,XMM}) = 0.9(\pm 0.03) \times log(L_{X,Chandra}) + 4.46(\pm 1.2)
\label{eq:hleq}
\end{equation}
If sources have been observed in the XMM survey, we adopt the hard X-ray luminosity derived from the XMM catalog. For the remaining sources in the Chandra catalog, their hard X-ray luminosity is converted to the XMM band system using Equation~(\ref{eq:hleq}). 
The new uncertainty of their hard X-ray luminosity is computed according to the 1$\sigma$ dispersion of the hard X-ray luminosity conversion equation. 
\emph{Correcting the systematic difference between the hard X-ray luminosities measured with Chandra and those measured with XMM, is the final step in ensuring that all hard X-ray luminosities are in the XMM band system.}

\begin{figure}
      \includegraphics[width=84mm]{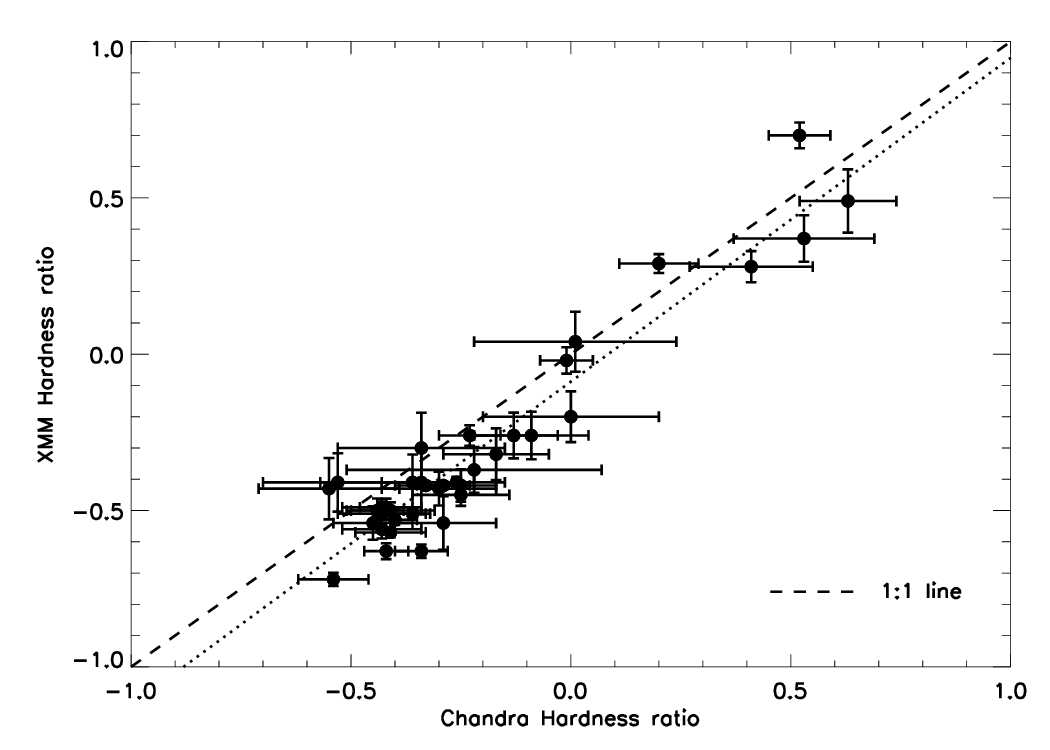}
         \caption{Hardness ratio extracted from the XMM catalog \citep{Brusa2010} vs. value from the Chandra catalog \citep{elvis2009} for our 38 sources detected in soft and hard bands with both observatories. The different hard X-ray energy bands between the two observations influence the hardness ratio measurements. We fit a simple line (dotted line) to estimate a conversion equation between the two different measurements (see Equation~(\ref{eq:hreq})). The black dash line indicates 1:1 line.}
\label{fig:hr}
\end{figure}

One of the main goals of our study is to explore the relationship between the host galaxy extinction and the central region obscuration. 
The X-ray hardness ratio (hereafter HR), defined as HR = (H - S)/(H + S), where H is the photon counts in the X-ray hard band and S is the photon counts in the soft band, is typically used as an indication of source spectral shape. Assuming a primary power-law spectrum, the neutral hydrogen column density can be measured by HR. Unlike the X-ray spectral fitting, which requires relatively good S/N, the HR can be used over a wide range of S/N.
Our HR values are extracted from \citet{Brusa2010} for XMM catalog, and \citet{elvis2009} for Chandra catalog. The difference in hard X-ray energy band between the two set of observations may change the HR values. Therefore, given that our samples overlap, we can fit a line and derive an empirical HR conversion equation. In Figure~\ref{fig:hr}, we compare the HR from XMM and Chandra, for the overlapping 38 sources detected in soft and hard bands with both observatories. 
The HR conversion equation is:
\begin{equation}
HR_{XMM} = 1.04(\pm 0.06) \times HR_{Chandra} - 0.09(\pm 0.02)
\label{eq:hreq}
\end{equation}
If sources have been observed in the XMM survey, we adopt the HR derived from XMM catalog. For the remaining sources with HR measurements in the Chandra catalog, these are converted to the XMM band system using Equation~(\ref{eq:hreq}). 
The new uncertainty of their HR is computed according to the 1$\sigma$ dispersion of the HR conversion equation. 
\emph{Correcting the systematic difference between the HRs measured with Chandra and those measured with XMM, ensures that all HRs are in the XMM band system.}
Then, we assume that there are no systematic differences between the HR measurements from both samples. We will later discuss the implication of such assumption in section~\ref{sec:hr-smx}.
It is worth to remind that time-dependent variability of AGNs (both photon index and flux) can induce stochastic inconsistency in different observations: for instance, \citet{Nateos2007} have summarized that a variation in slope of $\Delta$ $\Gamma$ $\sim$ 0.2 could lead to a variation in HR of $\Delta$ HR $\sim$ 0.1.

From our total sample of 142 \smx galaxies, 96 (68\%) are detected in both soft and hard X-ray bands. For the remaining 46 (32\%) objects, 24 are soft band detection, 21 are hard band detection, and 1 is full band detection (neither detected in soft band nor hard band).
We examine the overall $L_{\rm IR}$ and redshift distribution and find no significant difference between these 96 \smx galaxies with measured HRs and the rest of the sample, suggesting that there is no obvious bias within this subsample.

For the purpose of deriving unabsorbed hard X-ray luminosity, we need to correct the neutral hydrogen absorption. We use HR to calculate the column density of neutral hydrogen ($N_{H}$). 
One caveat of such method is that it relies on assumption of a simple absorbed power-law model. 
Extra components in the source spectra can affect HR measurements, including soft X-ray excess (at 1-2 keV), Fe K$\alpha$ line, and X-ray reflection, which are common in AGN and \ul\ \citep{Teng2010}.
These extra components may lead to an underestimation or even prevent to derive an intrinsic $N_{H}$.
Here, the low X-ray counts of each individual source prevents us to perform X-ray spectrum fitting directly. We therefore use a more uniform and simple procedure to derive a $N_{H}$ from the HR measurement.

\begin{figure}
      \includegraphics[width=84mm]{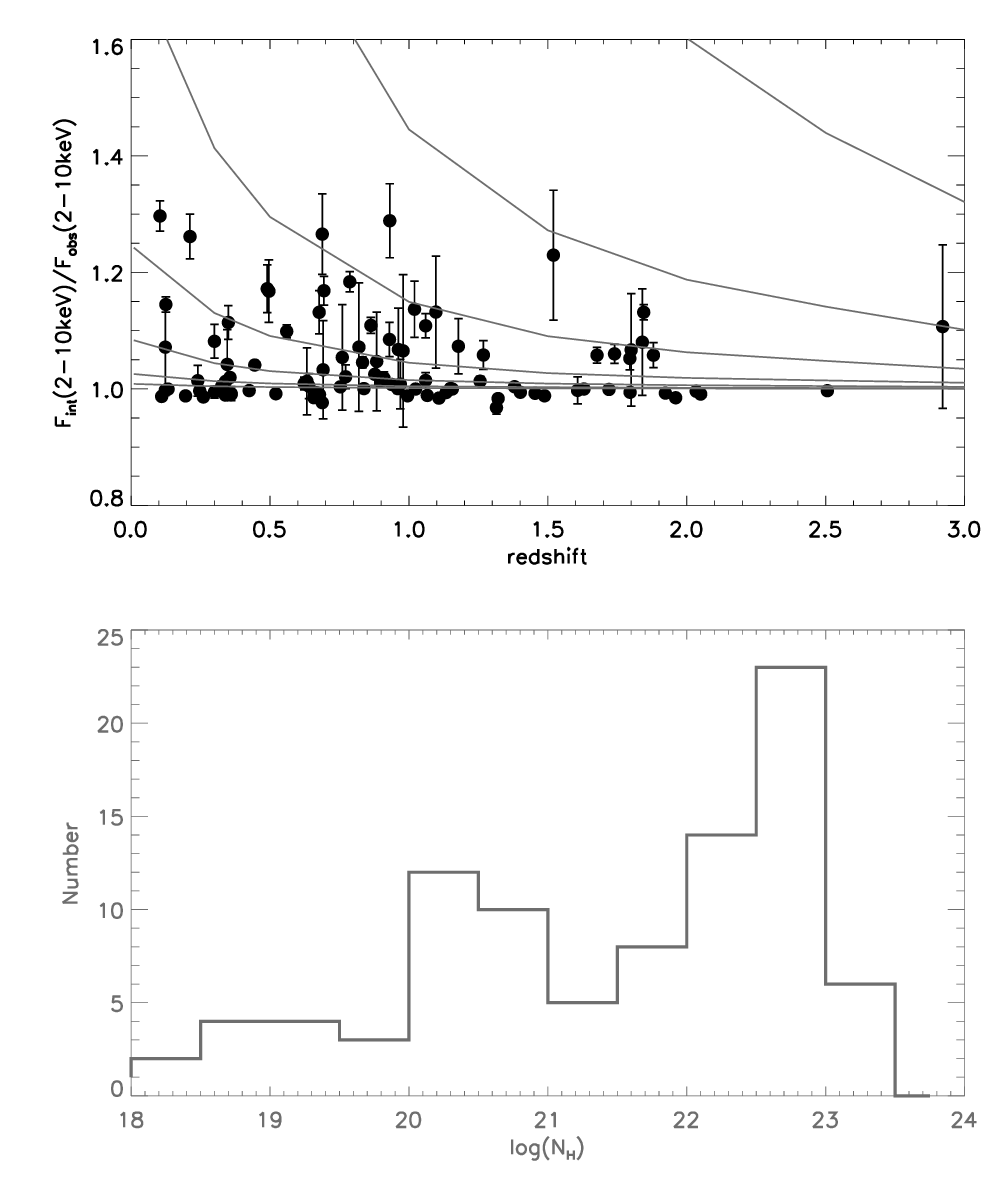}
         \caption{Top panel: The absorption correction factor F$_{int}$(2-10keV)/F$_{obs}$(2-10keV) versus redshift. F$_{obs}$ and F$_{int}$ indicate an observed 2-10 keV flux and an absorption corrected 2-10 keV flux in observed frame. Error bar comes from 1 $\sigma$ dispersion of HR. The solid lines from bottom to top present different amount of $N_{H}$ = 1 x 10$^{21}$, 3 x 10$^{21}$, 1 x 10$^{22}$, 3 x 10$^{22}$, 1 x 10$^{23}$, 3 x 10$^{23}$, and 1 x 10$^{24}$ cm$^{-2}$, respectively. Absorption correction factors with $< 1$ are set to be 1. Bottom panel: The histogram of column density of \smx galaxies.} 
\label{fig:frfo}
\end{figure}

First, we fix the intrinsic photon index of power-law to 1.7 \citep{Wilkes2005}, as a general broad-line AGN. 
To simulate the observed HR, we use the \textit{Portable, Interactive, Multi-Mission Simulator} (PIMMS)\footnote{http://heasarc.gsfc.nasa.gov/Tools/w3pimms.html \label{fn:pimms}} from HEASARC to model simulated sources with variable amount of $N_{H}$ covering a redshift interval 0 $<$ z $<$ 3. Then, by comparing the model HR to the observed HR within a given redshift, we could obtain our best $N_{H}$ value. 
The mean value of HR for \smx galaxies in COSMOS field is -0.16 with corresponding $N_{H}$ = 2.7 x 10$^{22}$ cm$^{-2}$ (at z = 0.95). 
The bottom panel of Figure~\ref{fig:frfo} shows the $N_{H}$ distribution of our \smx sample.
A caveat of estimating $N_{H}$ from the HR, is that the HR becomes less sensitive to absorption for high redshift galaxies (e.g. observed-frame 0.5-2 keV shifts to rest-frame 1.5-6 keV at z = 2), except for those very heavily obscured AGNs. Therefore, in this study, the $N_{H}$ of \smx galaxies at high redshift could be unavoidably underestimated.

Then, we apply the best-fitted value of $N_{H}$, redshift, and photon index $\Gamma \sim$ 1.7 to measure the absorption correction factor between absorbed and unabsorbed hard X-ray luminosity for \smx galaxy. 
The top panel of Figure~\ref{fig:frfo} shows the absorption correction factor as a function of redshift with mean value of F$_{int}$/F$_{obs}$ = 1.05. 
We denote an observed 2-10 keV flux as F$_{obs}$ and an absorption corrected 2-10 keV flux in the observed frame as F$_{int}$.

Finally, given the large redshift range covered by our sample (see Figure~\ref{fig:nz}), it is necessary to apply k-corrections to compute the X-ray rest-frame luminosities of our \smx galaxy sample,which we apply after the absorption correction.
We assume that the spectrum of our galaxies in X-ray follows a simple power-law with photon index $\Gamma$ = 1.7, and derive the ratio between rest-frame and observe-frame flux integration, to compute the rest-frame flux. Although using different $\Gamma$ values could affect the derived k-correction factor, such variation is not significant. 
We apply a correction based on $\Gamma$ = 1.7 to our \smx galaxies disregarding them being from the Chandra or XMM catalog.

To ensure the intrinsic 2-10keV luminosity robustness, we only estimate the absorption correction for those samples that have detection both in soft and hard band. 
It is worth noting that galaxies only detected in hard X-ray band (i.e. HR = 1) can also indicate the presence of Compton-thick AGNs ($N_{H}$ $\ge$ 1.5 x 10$^{24}$ cm$^{-2}$). Compton-thick AGNs are usually included in X-ray background synthesis model to interpret the intensity peak about 30 keV \citep{Comastri2004}.
We try to quantify the fraction of obscured AGN in our sample by binning hard X-ray luminosity in a range of 41 $\le$ log $L_{X}$ $<$ 43, 43 $\le$ log $L_{X}$ $<$ 45, and 45 $\le$ log $L_{X}$. For each of these hard X-ray luminosity bins, the fraction of obscured AGNs with log $N_{H}$ $\geq$ 22 to all AGNs (in the same hard X-ray luminosity range) is $\sim$ 44\%, 48\%, and 29\%, respectively. 
\citet{Ueda2003} presented the hard X-ray luminosity function and absorption function of AGN up to redshift = 3.
In Figure 7 of their paper, the obscured AGN fraction is 58$\pm$1\%, 50$\pm$3\%, and 34$\pm$5\% with corresponding hard X-ray luminosity range of 41.5 $\le$ log $L_{X}$ $<$ 43, 43 $\le$ log $L_{X}$ $<$ 44.5, and 44.5 $\le$ log $L_{X}$.
We compare the obscured AGN fraction of \smx galaxies with \citet{Ueda2003}.
The obscured AGN fraction of the higher X-ray luminosity (i.e. 43 $\le$ log $L_{X}$ $<$ 45 and 45 $\le$ log $L_{X}$) in our study is comparable to \citet{Ueda2003}. 
On the contrary, our lower X-ray luminosity (i.e. 41 $\le$ log $L_{X}$ $<$ 43) range, probably miss around 20\% of obscured AGN, implying that the sample with this range loses more obscured AGNs relative to our higher X-ray luminosity sample.
However, more recent studies inferred a different trend in the obscured AGN fraction than \citet{Ueda2003}, in particular with a non-negligible incompleteness detected for very high luminosities \citep[log $L_{X}$ $\ge$45 - ][]{Gilli2007,Merloni2014}. Fortunately, 70\% of our sample with intrinsic 2-10keV luminosity information, is in a range of 43 $\le$ log $L_{X}$ $<$ 45, where the obscured fraction is comparable with the literature \citep[e.g. ][]{Ueda2003,Gilli2007,Merloni2014,Ueda2014}.

\section{RESULTS}

\subsection{AGN versus star formation}
\label{sec:agn-dia-all}

Our sample is drawn from both $70\micron$ and X-ray catalogs, the far-infrared detection indicates the presence of dust from associated star formation. In fact, the star formation may generate the X-ray emission by supernovae remnant and high mass X-ray binaries (HMXBs) instead of being generated by a single powerful AGN.  
Therefore, in this section, we try to use different methods to study the driving mechanism of \smx galaxies.

\subsubsection{AGN criterion from colour-colour selection}
\label{sec:agn-ccsel}
The mid-infrared IRAC colour-colour selection described by \citet[hereafter S05]{2005ApJÉ631..163S} is widely accepted as a robust method to select AGN candidates. 
Such colour criteria are efficient to separate AGN candidates from star-forming galaxies due to the red $[3.6\micron]-[4.5\micron]$ colour inferred by the combination of the power-law continuum spectrum of the AGN and the relative weakness of the stellar bump feature at 1.6$\micron$ usually emitted by the host galaxy. \citet{Lacy2004} proposed another AGN diagnosis using a different IRAC colour-colour selection from S05. However, \citet{choi2011} used complete AGN samples based on different selections (e.g. BPT diagram, [O III]/H$\beta$ ratio, high excitation line [Ne V], and broad line region feature) and showed that the S05 colour-colour selection is more robust than \citet{Lacy2004}.

\begin{figure}
     \includegraphics[width=84mm]{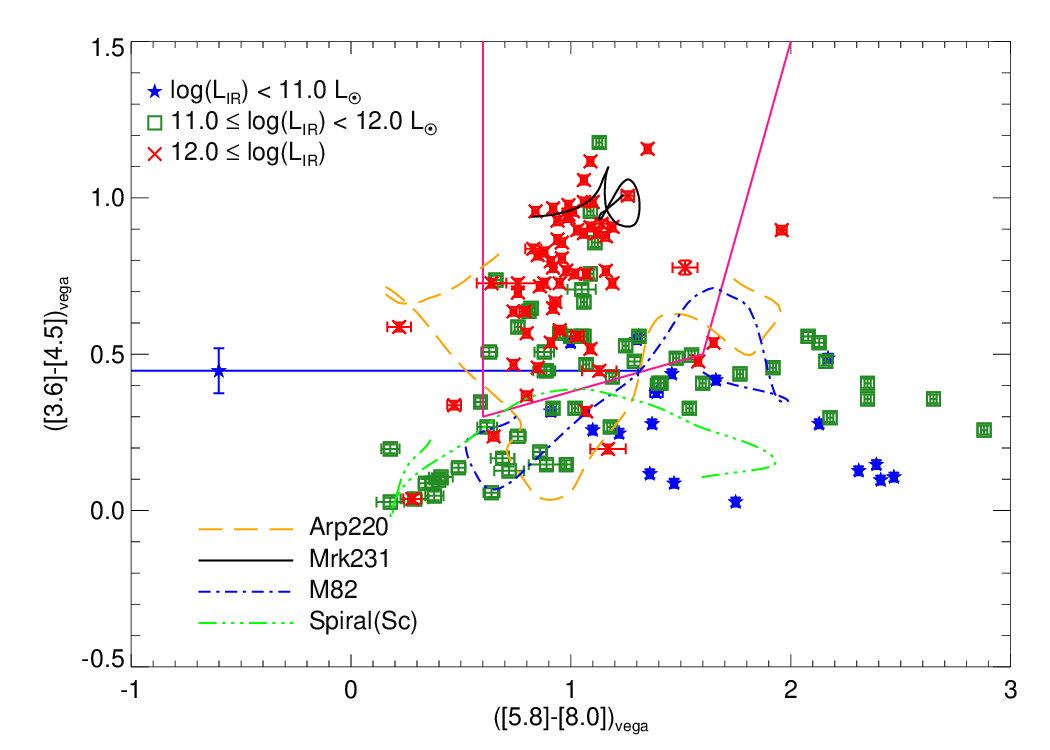}
        \caption[colour-colour diagram]{[3.6]-[4.5] versus [5.8]-[8.0] colour-colour diagram of \smx galaxies. The different $L_{\rm IR}$-selected galaxy samples follow the description in Figure~\ref{fig:nz}, with blue stars, green squares and red crosses indicating {\sfgs}, {\li}s and {\ul}s, respectively. The pink solid line represents the boundaries of the AGN region defined by \citet{2005ApJÉ631..163S}. Orange dash line and black solid line represent the evolution with redshift from z = 0 to 3 of the colours of star-forming dominated ULIRG Arp220 and AGN-dominated ULIRG Mrk231, respectively. Blue dash dot line and green dash-triple-dot line represent the evolution with redshift from z = 0 to 1 of the colours of starburst galaxy M82 and Sc-type spiral galaxy templates, respectively.}
\label{fig:agn-ccsel}
\end{figure}

In Figure~\ref{fig:agn-ccsel} we show the IRAC [3.6\micron]-[4.5\micron] vs. [5.8\micron]-[8.0\micron] colour-colour diagram of our \smx galaxies. The deep pink line indicates the AGN dominating region defined by S05. Using these criteria, we separate AGNs from our {\sfgs}, {\li s}, and {\ul s}. Their estimated AGN fractions are 20{\%} (4 out of 20; blue stars), 48{\%} (29 out of 60; green square), and 92{\%} (57 out of 62; red cross), respectively. There are 63{\%} (90 out of 142) of \smx galaxies classified as AGN from the S05 selection criteria.
We observe that {\ul}s are overlapping more with AGN region in the diagram than the locus of {\sfgs}; this result is consistent with literatures that demonstrate an increasing AGN fraction with increasing infrared luminosity \citep{France2003,Lee2010,Wang2010}.

Our sample spans a wide range of redshifts which may influence the observed colour-colour properties due to k-correction effect.
To better quantify this redshift effect, we place the SED of templates in different redshifts and simulate the colour changes in Figure~\ref{fig:agn-ccsel}.
Since 44\% of our \smx galaxies (62 out of 142) are classified as {\ul}s, we used the SEDs of the {\ul} Mrk231, which has a strong AGN contribution in the mid-infrared band, and of Arp220, which has intense starburst activity, and shift them to the redshift range of 0$<$z$<$3 (the {\ul}s in our sample span the redshift range of z = 0.5 to 3), then plot the colour changes.
Based on its IR properties, Arp220 is considered as the prototypical ULIRG powered by star formation. Despite the detection of high ionization Fe K lines in its X-ray spectrum preventing us from completely ruling out the possibility of a low luminosity or Compton-thick AGN \citep{Iwasawa2005}, we still assume, as is commonly done in the literature, that Arp220 represents a star-formation dominated {\ul}. On the other hand, in the ULIRG Mrk231, the mid-infrared continuum can be well fitted by a power-law spectrum \citep{Weedman2005} and the hard X-ray emission is dominated by an unresolved nuclear point source \citep{Gallagher2002}.
It is unambiguously accepted that an AGN hosts in the centre. We use Mrk231 to represent an AGN-dominated {\ul}.
In Figure~\ref{fig:agn-ccsel}, the Mrk231 colours remain in the S05 AGN region at any redshift because of its featureless power-law SED in mid-IR. In contrast, the Arp220 colours may satisfy the S05 AGN criteria in some redshift ranges because of the 3.3$\micron$, 6.2$\micron$, and 7.7$\micron$ PAH emission features entering into the IRAC photometry bands and confusing the AGN diagnosis. The S05 criteria are, therefore, not extremely robust against Arp220-like star formation dominating {\ul}.

However, 56{\%} of \smx galaxies (80 out of 142) are not classified as {\ul}s, their $L_{\rm IR}$ cannot compete with Mrk231 and Arp220. Therefore, we place the SED of starburst galaxy M82 and main-sequence Sc-type spiral galaxy in the redshift range of 0$<$z$<$1 ({\li}s and star-forming galaxies are below z = 1 in our sample) and simulate the colour changes. Figure~\ref{fig:agn-ccsel} shows that in most redshift intervals, both M82 and Sc-type spiral galaxy are almost in the outside of the AGN boundaries. 
All SED templates used above come from the SWIRE galaxy library, and are a combination of models, photometric data, and observed IR spectra \citep[and references therein]{Polletta2007}.

\subsubsection{$L_{\rm X}$ vs. $L_{\rm IR}$ relation}
\label{sec:lxlir}

The relation between hard X-ray luminosity and total infrared luminosity provides us with another method to determine whether star formation from host galaxy or AGN predominates the overall SED. While colour-colour selection can be altered by the presence of specific lines \citep{Donley2008}, total infrared luminosity is more robust because it is derived from 8-1000\micron\ continuum \citep{Smail2011}.
 
\begin{figure}  
  \includegraphics[width=84mm]{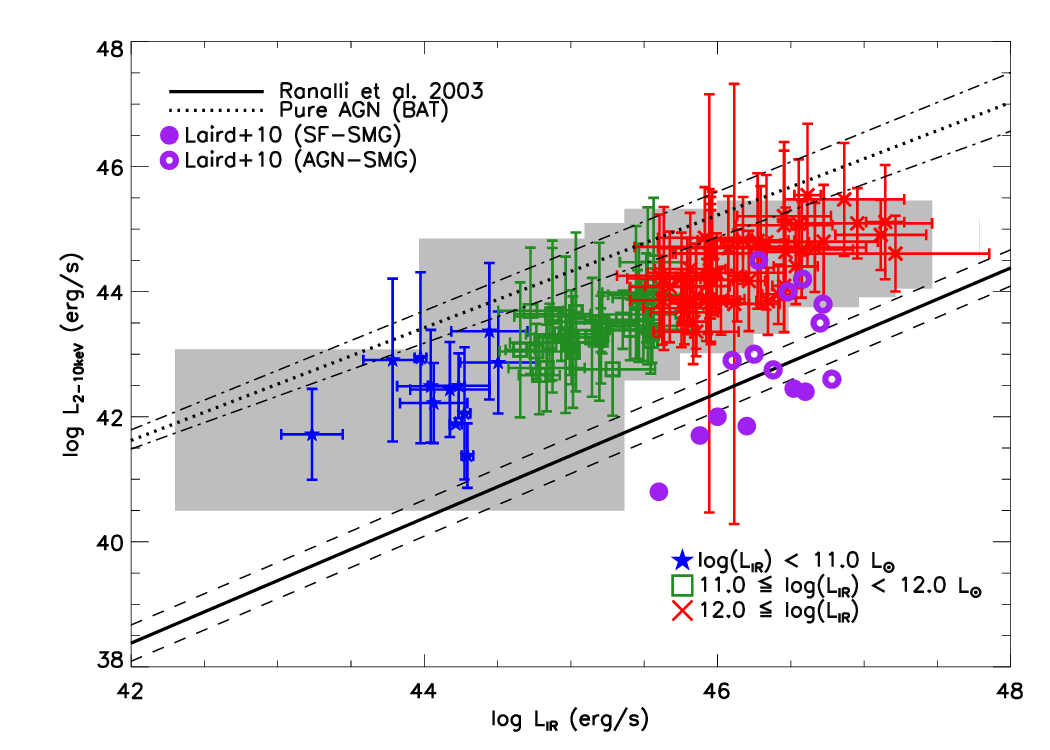}
        \caption[LX-LIR]{Absorption-corrected rest-frame 2-10keV luminosity versus total infrared luminosity of \smx galaxies. Symbols are the same as in Figure~\ref{fig:agn-ccsel}. The X-ray luminosity error bars are derived from the estimated flux errors, while the total infrared luminosity error bars are the 1$\sigma$ uncertainty based on the result of \citet{Kartaltepe2010} template fit. Grey regions represent the areas covered by maximum and minimum luminosities in different redshifts bins drawn from the full X-ray selected and full $70\micron$ selected samples. These observational limitations induce the apparent strong correlation between $L_{\rm X}$ and $L_{\rm IR}$ observed in COSMOS field. Black dot line and black solid line represent the $L_{\rm X}$ vs $L_{\rm IR}$ relations for pure AGNs \citep{Mullaney2011} and star-forming galaxies \citep{Ranalli2003}, respectively. Dash-dot line and dash line represent 1$\sigma$ dispersion of the $L_{\rm X}$ vs $L_{\rm IR}$ relations for pure AGNs and star-forming galaxies. \citet{Laird2010} separated the star forming submillimeter galaxies (SF-SMGs) and the AGN-hosted submillimeter galaxies (AGN-SMGs) from X-ray spectrum fitting. Fill and open circles represent SF-SMGs and AGN-SMGs, respectively.}
     \label{fig:lxlir}
\end{figure}

In Figure~\ref{fig:lxlir} we display the absorption-corrected rest-frame hard X-ray ($2-10$keV) luminosity against k-corrected total ($8-1000\micron$) infrared luminosity for our \smx galaxies, which we divide into three subsamples according to their $L_{\rm IR}$ luminosity: \sfgs, \li s, and \ul s (same symbols as in Figure~\ref{fig:agn-ccsel}).
$N_{H}$ absorption correction and k-correction have been applied to the hard X-ray luminosity for those 96 sources with HR measurement (please see Section~\ref{sec:hr} for details). For the remaining 46 sources without HR measurement, even if we have redshift information to perform a k-correction of X-ray luminosity, the lack of HR measurement prevents us from performing an $N_{H}$ absorption correction, and therefore they are not included in Figure~\ref{fig:lxlir}.
The X-ray luminosity versus infrared luminosity plot shows a tight Spearman's rank correlation coefficient, $\rho$ $\sim 0.88$. (Note that SpearmanÕs $\rho$ = 1 corresponds to two variables being monotonically related.)
However, this strong correlation may be a manifestation of an observational bias. In order to understand this tight correlation, and whether it comes from the observational bias, we need to estimate the upper and lower limits in the $L_{\rm X}$ vs. $L_{\rm IR}$ plot. 
For the total infrared luminosity, we slice the whole $70\micron$ catalog \citep{Kartaltepe2010} into several redshift bins from z = 0.033 to z = 3, and identify the maximum and minimum total infrared luminosity.
For the hard X-ray luminosity, we simulate the lower X-ray luminosity in the same redshift bins using the Chandra flux limits and measure the upper X-ray luminosity directly from  the XMM catalog. 
The results are displayed as grey regions in Figure~\ref{fig:lxlir}.
These observational limits and their evolution with redshift indeed explain the tight correlation between the infrared and X-ray luminosities. 

By studying X-ray properties of \sfgs, \citet{Ranalli2003} demonstrated that the hard X-ray luminosity is also proportional to the star formation rate (SFR). This relation between $L_{\rm X}$ and $L_{\rm IR}$ is represented by the black solid line in Figure~\ref{fig:lxlir}. In fact, the most important contributors of X-ray emission in such galaxies are the high mass X-ray binaries (HMXBs). \citet{France2003} explored the X-ray properties of a sample of {\ul}s and most of them do not exhibit AGN signature. The conclusion of their work is that the X-ray luminosity and spectral shape of some {\ul}s are dominated by hot thermal plasma and X-ray binaries originating in recent starburst region.

On the other hand, because of the presence of the parsec-scale dust ``torus" surrounding the accretion disk of the central SMBHs, AGNs emit light in the infrared wavelength \citep{Gandhi2009}. Indeed, the ultraviolet and optical light emitted by the central accretion disk is absorbed by the dust and re-emitted in infrared. 
\citet{Gandhi2009} observed the core of nearby AGNs with unprecedentedly high spatial resolutions in both mid-IR and X-ray wavelength, demonstrating a strong correlation between $12.3\micron$ and hard X-ray luminosities. Although, the rest-frame at 12.3$\micron$ could be influenced by the PAH emission feature at 11.3$\micron$, the dispersion of $L_{\rm X}$/$L_{\rm 12.3\micron}$ is quite small which may be due to statistical uncertainty rather than physical reasons \citep{Lutz2004, Gandhi2009}.
Based on the intrinsic AGN/quasar IR SED from \citet{Netzer2007}, \citet{Mullaney2011} established an empirical relation between the $12.3\micron$ luminosity and the AGN infrared luminosity (see Figure 11 of their paper). 
By combining with the relation between $12.3\micron$ and hard X-ray luminosities established by \citet{Gandhi2009}, \citet{Mullaney2011} derived an unbiased $L_{\rm X}$ vs. $L_{\rm IR}$ relation for AGNs  (see equation (4) of their paper), free of any contamination from the host galaxy.
This ``pure'' AGN $L_{\rm X}$ vs. $L_{\rm IR}$ relation is represented by the dotted line in Figure~\ref{fig:lxlir}.  
This relation is confirmed observationally by combining the swift-BAT X-ray selected AGN population \citep{Tueller2010}, together with IRAS infrared measurements (BAT/IRAS AGN). The strong $60\micron$ luminosity of this population is more likely powered by the AGN rather than star formation activity from the host galaxy, as demonstrated by the linearly increasing correlation between $L_{\rm IR}$ and $L_{\rm X}$ \citep{Mullaney2012}. In addition the hard energy band of swift-BAT observation (14-195 keV) enables a selection of galaxies unbiased against obscuration. 

As shown in Figure~\ref{fig:lxlir}, the $L_{\rm X}$ and $L_{\rm IR}$ distribution of our \smx galaxies deviates from these established "pure" AGN and "pure" star forming relation by $0.5-1$ dex and $1-2$ dex, respectively. 
Ignoring the pure-AGN and pure-SF relations are affected by any observational bias, our sample indicates a higher AGN activity relative to star formation.
An important note of caution is that this result is strongly dependent of observational limits, as shown by the grey region of Figure~\ref{fig:lxlir}. Because shallow X-ray observations in COSMOS are limited in the detection of "pure" star forming galaxies, especially at high redshift, our sample is biased toward most sources with higher X-ray emission, likely to host an AGN.
On the other hand, for a given $L_{\rm X}$, a higher $L_{\rm IR}$ may indicate a contamination from star formation in the host galaxy, shifting our sample away from a ÒpureÓ AGN relation in the $L_{\rm X}$ vs $L_{\rm IR}$ diagram.

It is also interesting to compare the properties of our sample with a sample of Submillimeter galaxies (SMGs); a population of objects selected according to their detection in submillimeter wavelengths. This population is dominated by strongly star-forming cold-dust enshrouded galaxies at high redshift \citep*[e.g.][]{Chapman2005}.
The concomitance in redshifts and total infrared luminosities with \ul s may indicate that SMGs are an early stage of evolution of the merger scenario \citep{Greve2005,Biggs2008,Engel2010}. 
Ultra-deep X-ray observations of those distant star-forming galaxies indicate that 20-30\% of SMGs host an AGN \citep{dma2005}.
\citet{Laird2010} have studied the X-ray spectral properties of SMGs and classified them as AGN or starburst. These objects are represented in Figure~\ref{fig:lxlir} with open and closed purple circles, respectively. Our \smx galaxies appear to share similar $L_{\rm X}$ vs. $L_{\rm IR}$ relation as the AGN-hosted SMGs.

In summary, all indicators in the $L_{\rm X}$ vs. $L_{\rm IR}$ distribution of our \smx galaxies suggest that there is a significant contribution from AGN activity in the X-ray band relative to pure star formation.



\subsection{Hardness ratio of \smx galaxies}
\label{sec:hr-smx}

\begin{figure}
      \includegraphics[width=84mm]{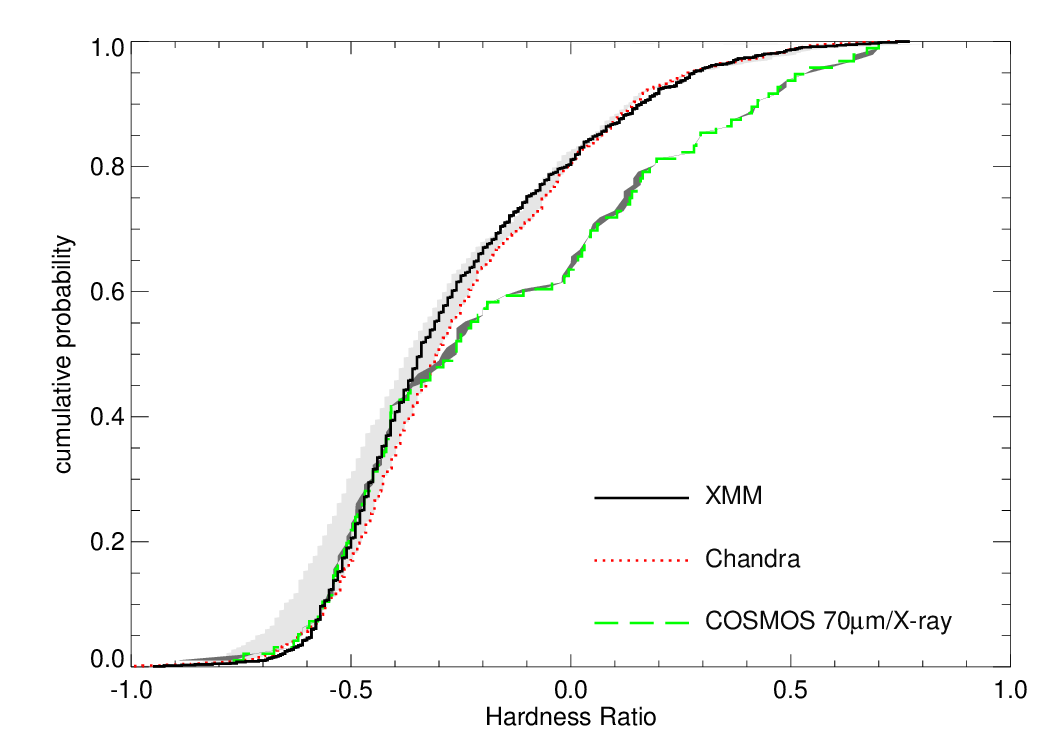}  
         \caption{Kolmogorov-Smirnov test: the hardness ratio cumulative probability distribution of XMM, Chandra, and \smx selected samples are represented in black solid, red dotted, and green dash colour lines, respectively. The sample number of XMM, Chandra, and \smx selected samples are 1001, 920, and 96. The light grey and dark grey regions represent the 1-$\sigma$ uncertainties in HR conversion equation of Chandra and \smx sample. The mean values (1 $\sigma$) of hardness ratio of XMM selected sample, Chandra selected sample, and \smx sample are -0.26 ($\pm$ 0.28), -0.28($\pm$0.29), and -0.16($\pm$0.39), respectively. Distribution of XMM and Chandra indicate these samples are drawn from the same population. However, the distribution of \smx sample and X-ray selected samples seem not to be drawn from the same population. The distributions indicate an excess of HR $\ge$ -0.3, revealing that the cold dust from the host galaxy may be responsible for additional obscuration.}
\label{fig:ks}
\end{figure}

To explain the diversity of AGNs (e.g., narrow-lines vs. broad-lines), an unified model based on the variation of obscuration due to the orientation of the dust torus surrounding the SMBH accretion disk, has been proposed. In such model, the presence of an edge-on dust torus not only blocks the broad-line emissions, but also increases the absorption in soft X-ray because of the higher hydrogen column density \citep{Antonucci1993}.

In parallel, galaxy mergers can accumulate large amount of neutral hydrogen and also trigger star formation \citep[e.g.,][]{Esquej2012}. 
Using the Extended Chandra Deep Field-South (ECDF-S) dataset, \citet{Treister2009} showed that AGNs residing in star-forming galaxy display large absorption in soft X-ray emission, indicating a large amount of neutral hydrogen existing along the direction of line-of-sight.

Our aim through this work is to investigate whether there is a direct connection between the obscuration properties of the AGN and those of the host galaxy.
As absorption affects the soft X-ray emissions more than the hard X-ray emissions, the HR is a good tracer of obscuration, with higher value of HR corresponding to larger absorption.
We apply Kolmogorov-Smirnov (K-S) test on two datasets, X-ray selected galaxies and \smx galaxies, to address their likelihood to be drawn from the same sample (i.e. our null hypothesis) \citep[{\it Numerical Recipes}, ][]{Press1992}.

As an initial test, we apply the K-S test on the HR measurements of both XMM and Chandra X-ray selected sources to test the robustness of our conversion between the two systems (for details, see section \ref{sec:hr}). The results of our K-S test for these populations are shown in Figure~\ref{fig:ks}, with the HR cumulative probability distribution of XMM selected sample in black and Chandra selected sample in red, and the K-S parameters are summarized in Table~\ref{tab:ks}. The probability of K-S test for XMM and Chandra selected sample to be drawn from the same population is only 3\%, which is not significant enough to reject the null hypothesis (our threshold to reject the null hypothesis is 99\%; in other words, a probability smaller than 0.01 rejects the null hypothesis). 

The size of the XMM and Chandra selected samples is 1001 and 920 sources, respectively, which is larger than the 96 sources included in our \smx galaxy sample. 
In order to better understand how sample size impacts the results of our K-S test, we proceed in a simple random sampling method. We randomly select 96 sources out of the XMM (Chandra) selected sample and run a K-S test between this subsample and the Chandra (XMM) selected sample. 
This procedure has been repeated for 1000 times to estimate probability mean and standard deviation. Following this procedure, we estimate that the probability of K-S test for XMM and Chandra selected sample to be drawn from the same population can be as high as $\sim$ 40\% when including the sample size errors; these results are summarized in rows 2 and 3 of Table~\ref{tab:ks}. 
Although two X-ray facilities have different sky coverage and depth, both XMM and Chandra HR distribution are roughly identical. 

Thereafter, we apply the K-S test on the \smx and XMM selected samples. This time, the null hypothesis is rejected at a significance of 99\% confidence level, implying that \smx galaxies and XMM galaxies are drawn from different populations. The K-S test on \smx and Chandra selected samples also rejects the hypothesis. The probability for both samples to be drawn from the same population is 0.3\% for XMM \& \smx sample, and 0.2\% for Chandra \& \smx sample. Figure~\ref{fig:ks} displays the cumulative probability distribution of HR from XMM (black solid line), Chandra (red dotted line), and \smx (green dash line) selected samples. The light grey and grey regions represent the effect on the cumulative distribution of HR of a 1-$\sigma$ uncertainty implied by our HR conversion equation on the Chandra and \smx samples (as discussed in section \ref{sec:hr}). Even when the HR conversion equation applied on the Chandra sample varies within a 1-$\sigma$ uncertainty (see Equation~(\ref{eq:hreq})), the cumulative distribution of both XMM and Chandra HR ratios are similar and the cumulative distribution of \smx HR is consistent with this sample not being drawn from the same population than XMM/Chandra purely X-ray selected samples.
The K-S parameters resulting from the three tests are summarized in Table~\ref{tab:ks}. 
It is useful to note that in above analysis the \smx detected sources are included in the XMM and Chandra selected samples. But even if we exclude the \smx detected sources from the XMM and Chandra selected samples, the main results of the K-S tests do not change.

In fact, the cumulative probability distribution below HR $\sim$ -0.3 is identical for the \smx galaxies and pure X-ray selected samples (either from XMM or Chandra observation), indicating that the fraction of unobscured AGNs is uniform over the full sample.
On the opposite, the obvious diversity of cumulative probability distribution appears above HR $\sim$ -0.3: there are 80\% of AGNs with HR $\leq$ 0 in the pure X-ray selected sample, while there are only 60\% of AGNs in the \smx galaxy sample. According to column densities we derived from the HR measurements, the mean value of $N_{H}$ for the overall \smx galaxy sample is $\sim$ 10$^{22}$ cm$^{-2}$, which is consistent with a mildly absorbed AGN population.
However, the \smx sample contains 21 galaxies detected only in the hard X-ray which are potential Compton-thick AGN (21/142; 15\%). These are obviously not included in our K-S tests because of lack of HR measurements, and they can potentially increase the difference between \smx galaxies and pure X-ray selected samples. 
It may be useful to mention that, for XMM and Chandra selected sample, the fraction of the sources that only have hard X-ray detection is 12\% (213/1797) and 15\% (257/1761), respectively. These fractions are consistent with those found for the \smx galaxy sample.
In order to estimate the column density for 21 hard X-ray detected \smx galaxies, given a known redshift, we can constrain their soft X-ray count rate from the flux limit of the observations and then calculate the lower limit of HR. Following the same procedure we used for \smx galaxies with HRs measurements (see Section~\ref{sec:hr}), we obtain a mean value of column density of at least $\sim$ 10$^{23}$ cm$^{-2}$ for those 21 hard X-ray detected \smx galaxies. 
Among these 21 hard X-ray detected \smx galaxies, one object (70$\micron$ ID = 1539) has high probability to be a Compton-thick AGN with the lower limit of HR $\sim$ 0.921, corresponding to $N_{H}$ $\sim$ 1.45 x 10$^{24}$ cm$^{-2}$.
With its photometric redshift (z=2.36), the derived total infrared luminosity may exceed 10$^{13}$ L$\sun$, putting this object in the 'hyper ULIRG' category. 
The exclusion of such objects from our analyses only reinforce the conclusions of our K-S tests that \smx galaxies are not from the same population as the full X-ray selected samples.
On the other hand, there are 24 soft X-ray detected \smx galaxies not included in our K-S tests because of lack of HR measurements, which can potentially influence the results. 
We adopt the hard X-ray count rate from flux limit of the observation and derive the upper limit of hardness ratio. 
A corresponding mean value of column density for soft X-ray detected \smx galaxies is lower than $\sim$ 10$^{20}$ cm$^{-2}$. 
It is important to note that the main results of K-S tests presented in this section does not change qualitatively if we included the soft X-ray detected \smx galaxies.


Our tests lead to the conclusion that the \smx galaxies include more obscured AGNs than pure X-ray selected samples.  
This result implies an additional process responsible for an increased $N_{H}$ in the line-of-sight of AGNs hosted by dust-enshrouded galaxies; process not accounted for in the conventional AGN unified model scheme. We speculate that the excess of obscuration is not only attributed to absorption on the line-of-sight by continuum or clumpy dust torus around the SMBH, but also is associated with an additional more diffused dust component generated by strong star formation in the host galaxy. In this case, the starburst could be the consequence of an event involving the whole system (e.g. merger) or generated by the circumnuclear region (e.g. bar). 
\citet{Kartaltepe2010b} have identified that the merger fraction (either major mergers or minor mergers) and AGN fraction increase systematically with increasing infrared luminosity (especially \ul s) in 70$\micron$-selected galaxies. The excess of obscured AGNs in our study could support the merger scenario; an intense starburst triggered by mergers produces large amount of dust and gas, which tends to hide the central AGN.

\begin{table*}
  \caption{Results of the K-S test on the hardness ratio between different catalogs. The randomly selected subsample are marked with an asterisk. Their mean value and standard deviation are computed with 1000 time random selection.}
    \begin{tabular}{llllllll} \hline               
  Sample 1     &   Number     &   Sample 2   & Number  &$D_{\rm max}$ & Significant  & Hypothesis \\ 
   &(sample 1) & &(sample 2) & & &(p=0.01) &  \\ \hline \hline
  XMM  & 1001 &Chandra    & 920   & 0.0664   & 0.0279  & Not Reject    \\ \hline
  XMM & 1001  &Chandra (Subsample)* & 96   & 0.100$\pm$0.026  & 0.397$\pm$0.261 & Not Reject \\ 
  Chandra  & 920  &XMM (Subsample)* & 96   & 0.0967$\pm$0.022  & 0.423$\pm$0.238  & Not Reject   \\ \hline
  XMM & 1001  & \smx  & 96   & 0.1920  & 0.0026  & Reject     \\
  Chandra  & 920 & \smx   & 96   & 0.1981  & 0.0018  & Reject    \\ \hline
    \end{tabular}
    \label{tab:ks}
\end{table*}

\subsection{Far-infrared colour of host galaxy}
 \label{sec:dust}

According to the predictions from the merger scenario, galaxy interaction triggers massive star formation, while at the same time, it disturbs the gas distribution, funneling it toward the central black hole, feeding the accretion disk, and therefore, generating a phase of nuclear activity \citep{Imanishi2010}. If the center of the galaxy harbors an AGN, dust in the host galaxy could be influenced by the radiations produced by the nucleus. 

In order to determine the temperature of cold dust component, ideally one would like to measure fluxes in the Rayleigh-Jeans tail of the blackbody, at longer wavelength such as 160$\micron$. 
Far-infrared photometry is an indicator of the dust properties of the galaxy.
In our \smx galaxies sample, there are 52 galaxies that have 160\micron\ detection (52 out of 142; 37{\%}). 
This fraction is comparable to that of the 160$\micron$-detected 70$\micron$ sources (463 out of 1503; 31{\%}) with or without X-ray detection.
Measuring colours at longer wavelength is a method to estimate the cold dust temperature \citep{Casey2012}. 
To explore whether the cold dust temperatures of \smx and 70$\micron$ galaxies differ, we use the observed frame colour in log($F_{70\micron}/F_{160\micron}$) as a proxy of the SED in far-infrared. A caveat of this method is that the rest-frame far-infrared photometry corresponding to the observed photometry shifts toward shorter wavelength as the redshifts of the galaxies increase, limiting our ability to reliably measure the cold dust temperature at higher redshifts.

\begin{figure}
      \includegraphics[width=84mm]{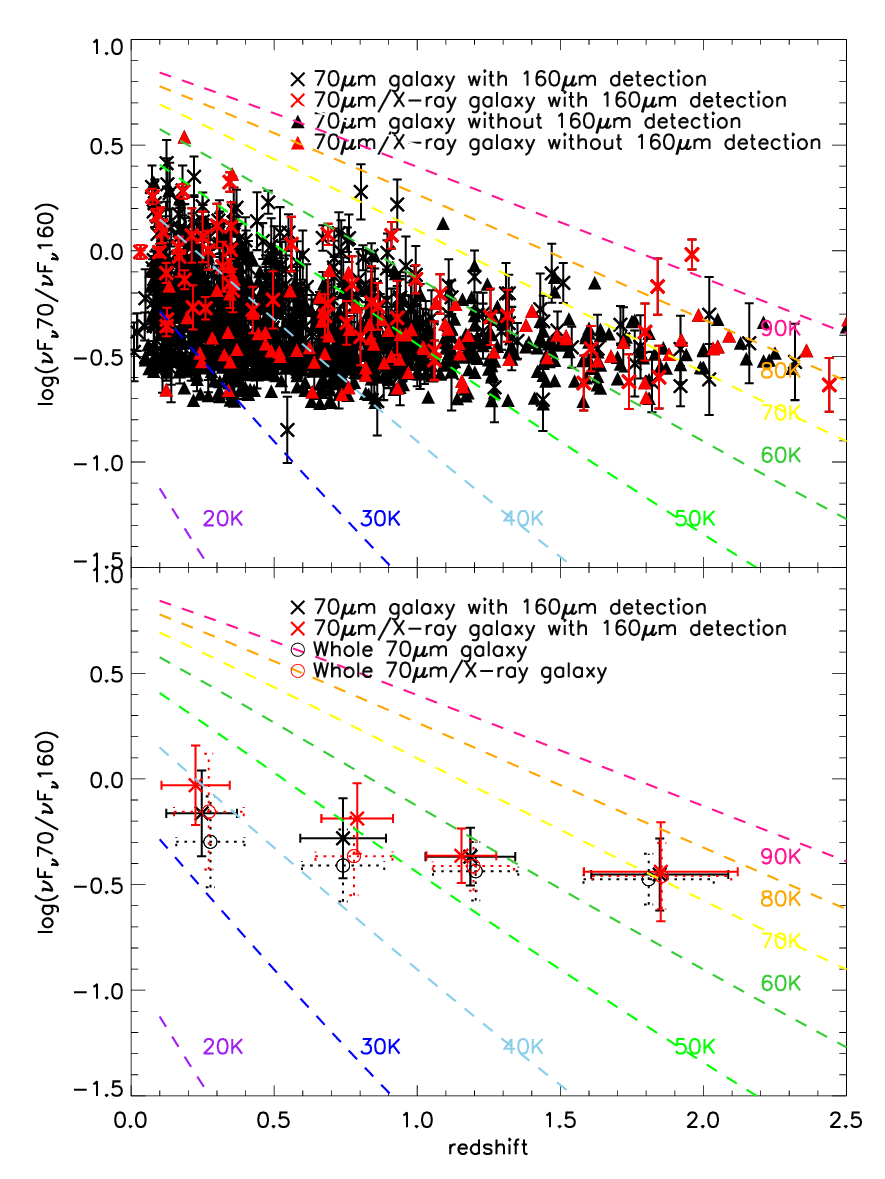}  
        \caption[Temperature]{ Top panel: observed frame log($F_{70\micron}/F_{160\micron}$) as a function of redshift; black and red symbols refer to all $70\micron$ galaxies and \smx galaxies, respectively. 70$\micron$ galaxies with and without 160$\micron$ detection are labeled as crosses and triangles, respectively. Dash lines in different colours show the simulated log($F_{70\micron}/F_{160\micron}$) as a function of redshift in different temperatures, derived from modified blackbody model with a given temperature. Bottom panel: Average log($F_{70\micron}/F_{160\micron}$) in observed frame versus redshift. Labeled colours and lines are same to top panel. Crosses represent the mean log($F_{70\micron}/F_{160\micron}$) of 70$\micron$ galaxies (or \smx galaxies) with 160$\micron$ detection. Open circles refer to the mean log($F_{70\micron}/F_{160\micron}$) of whole 70$\micron$ galaxies (or whole \smx galaxies), including those which only have lower limits.
 The error bars are the standard deviation of colour and redshift.} 

     \label{fig:temp} 
\end{figure}

The top panel of Figure~\ref{fig:temp} plots the log($F_{70\micron}/F_{160\micron}$) colour distribution of 70$\micron$ galaxies (black colour) and \smx galaxies (red colour) as a function of redshift. 70$\micron$ galaxies with 160$\micron$ detection are labeled as crosses, while 70$\micron$ galaxies without 160 detection are labeled as triangles. 
For those 70$\micron$ galaxies without 160$\micron$ detection, we set 50mJy as limiting flux density to measure lower limits of log($F_{70\micron}/F_{160\micron}$). 
A limit of 50mJy  results from iterating between the source extraction at different limiting flux densities and the confusion noise measurements until they converge to a ratio of 5; details are described in Sec. 5.2 of \citet{Frayer2009}.
Dash lines with varied colours represent the simulated log($F_{70\micron}/F_{160\micron}$) colours for different dust temperatures. When simulating the far-infrared colour, we only assume one cold dust temperature component. 
The synthetic MIPS photometry of 70$\micron$ and 160$\micron$ are obtained by performing the IDL routine, which assigns a temperature as an input blackbody spectrum to simulate photometric flux density (refer to Spitzer Data Analysis Cookbook \footnote{http://irsa.ipac.caltech.edu/data/SPITZER/docs/dataanalysistools/cookbook/ \label{ft:spitzer_idl}}, Recipe. 9). We slightly adjust this blackbody model by adding an emissivity parameter $\beta$, represented as an modified blackbody model: 
\begin{eqnarray}
{\nu}F_{\nu,model} = \nu^{\beta} B_\nu(\nu,T)
\label{eq:bb}
\end{eqnarray}
with a fixed emissivity $\beta \sim 1.5$ \citep{Clements2010}.
To understand the far-infrared colour evolving with redshift, we shift filters to rest-frame wavelengths and then measure the simulated log($F_{70\micron}/F_{160\micron}$) colour with a given redshift. We assign the temperature, ranging from 30K to 90K, typical dust temperatures for submillimeter galaxies (SMGs) and \ul s \citep{Chapman2005,Kov2006,Yang2007,Casey2009}.

In the bottom panel of Figure~\ref{fig:temp}, the mean log($F_{70\micron}/F_{160\micron}$) colour of our X-ray detected and not X-ray-detected 70$\micron$ galaxies are represented by red and black colours, respectively. The error bars are estimated from the standard deviation of log($F_{70\micron}/F_{160\micron}$) and redshift. 
Crosses and open circles represent the mean log($F_{70\micron}/F_{160\micron}$) colour of 70$\micron$ galaxies with and without 160$\micron$ detection, respectively. 
For the sources without a 160$\micron$ detection,  we simply set 50mJy as the 160$\micron$ limiting flux density as previously stated, and measure the lower limits of log(F70$\micron$/F160$\micron$). Given the large uncertainties, the average far-infrared colours for the 70$\micron$ galaxies and the \smx galaxies are similar in the same redshift interval. We do not see any significant difference in far-infrared colour between these two samples.

\section{DISSUSSION}
\subsection{AGN influence on the host galaxy dust temperature?}
 \label{sec:inf_dust}

Our analysis in section~\ref{sec:dust} shows that there is no difference in the far-infrared colour (i.e. log($F_{70\micron}/F_{160\micron}$)) of 70$\micron$ and \smx galaxy samples at all redshift. 
This result is in agreement with \citet{elbaz2010}, who derived an accurate estimation of the cold dust temperature of the host galaxy by using the Herschel photometry in the $250\micron$, $350\micron$, and $500\micron$ bands on the Rayleigh-Jeans portion of the spectra.
Their precise measurements show that, given the large error bars, the cold dust temperatures of galaxies with and without AGN are consistent.
In fact that there is no difference in far-infrared colours can be attributed to the different spatial scales between AGN and star formation: the cold dust is distributed across the several kpc scale of the galaxy and the energy released by the AGN may be insufficient to heat the dust at such scale. This implies that the mechanism of FIR emission in galaxies is mainly due to star formation disregarding the presence of an AGN.

\begin{figure}
      \includegraphics[width=84mm]{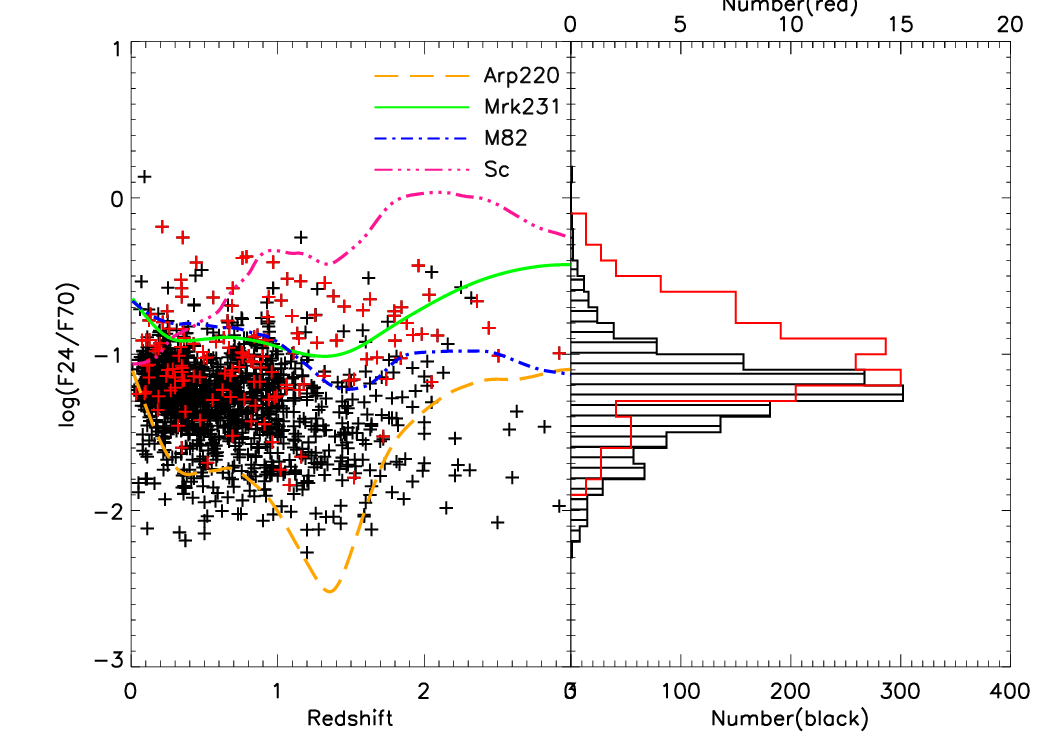}      
       \caption{Distribution of log($F_{24\micron}/F_{70\micron}$) colour index for $70\micron$ selected sample and \smx sample. $Left$ $panel:$  Distribution of the colour against redshift for the $70\micron$ galaxies (black crosses) and the \smx sample (red crosses). The green solid line, orange dash line, blue dash dot line, and pink dash dot dot dot line indicate the evolution of Mrk231, Arp220, M82, and normal spiral galaxy flux ratio as a function of redshift by using the templates from \citet{Polletta2007}.  $Right$ $panel:$ Histogram of the colour index for $70\micron$-selected galaxies is black and \smx is red. The red histogram has been normalized to the peak of black sample for a better illustration (bottom and top axis represent the number of $70\micron$-selected galaxies and \smx, respectively). The $70\micron$ galaxies have a median colour index of log($F_{24\micron}/F_{70\micron}$) $\sim$ -1.27$\pm$0.28, while the \smx galaxies have log($F_{24\micron}/F_{70\micron}$) $\sim$ -0.99$\pm$0.3. Both satisfy the local cold dust definition of log($F_{24\micron}/F_{70\micron}$) $\leq -0.7$ \citep{1985Natur, Sanders1988-2}. }
  \label{fig:2470}
\end{figure}

Using infrared colour log($F_{24\micron}/F_{70\micron}$) to sample the blackbody temperature profile, \citet{Rafferty2011} concluded that galaxies hosting an AGN present a warm dust temperature. We reproduce their analysis for our sample, as shown in Figure~\ref{fig:2470}. 
The $70\micron$ galaxies have a median colour index of log($F_{24\micron}/F_{70\micron}$) $\sim$ -1.27$\pm$0.28, while the \smx galaxies have log($F_{24\micron}/F_{70\micron}$) $\sim$ -0.99$\pm$0.30. We apply the K-S test for these two samples; the probability to be drawn from the same population is $\ll$0.01\%, indicating the difference between two log($F_{24\micron}/F_{70\micron}$) colour distributions in Figure~\ref{fig:2470} is statistically significant.
The \smx galaxies present a higher colour index on average than overall 70$\micron$ galaxies at all redshifts, suggesting the presence of a warmer dust component, a similar conclusion to \citet{Rafferty2011}. 
In fact, the 24$\micron$ band is sensitive to warm dust around AGN, explaining the discrepancies in log($F_{24\micron}/F_{70\micron}$) colour between X-ray detected and undetected 70$\micron$ galaxies.
It is noted that log($F_{24\micron}/F_{70\micron}$) and log($F_{70\micron}/F_{160\micron}$) trace different dust temperature components. 

To explain the difference in colour index, we look at the evolution of the log($F_{24\micron}/F_{70\micron}$) colour as a function of the redshift for four templates, the star-forming \ul\ Arp220, the AGN \ul\ Mrk231, the starburst galaxy M82, and the Sc-type spiral galaxy. 
Templates were described in Section~\ref{sec:agn-ccsel}. In the case of Arp220, the rest-frame $9.7\micron$ absorption feature is shifted into the $24\micron$ band at redshift $z \sim1.5$ causing an apparent lower value of the colour index. Similarly, at $z > 2$ the 7.7 and 8.2 $\micron$ PAH emission features enter the $24\micron$ band, raising the $F_{24\micron}/F_{70\micron}$ flux ratio. In the case of Mrk231, the AGN driven power-law continuum produces colour index with little variation for various redshifts. 
In the case of M82, its log($F_{24\micron}/F_{70\micron}$) colour is similar to Mrk231 but with lower $L_{\rm IR}$. 
However, M82 can only explain the log($F_{24\micron}/F_{70\micron}$) colour of star forming galaxies at z $<$ 0.5, as such objects are rare at high redshift in our sample. 
In the case of the main-sequence Sc-type spiral galaxy, because of the weakest infrared luminosity, the log($F_{24\micron}/F_{70\micron}$) colour is bluer than the other three templates.

In Section~\ref{sec:lxlir}, we show the possibility of higher AGN activity relative to star formation in \smx galaxies. This result is supported by the higher mean value of log($F_{24\micron}/F_{70\micron}$) in Figure~\ref{fig:2470}, in which the 24$\micron$ of \smx galaxies are powered by AGN. The \smx galaxies follow the similar log($F_{24\micron}/F_{70\micron}$) colours of the X-ray luminous galaxy, Mrk 231, across the redshift range $0 < z < 3$; while for most 70$\micron$ galaxies, they follow the values between Arp220-like and M82-like galaxies.

\subsection{AGN obscuration from starburst?}

According to the major merger scenario of galaxy evolution, developed to interpret a potential ULIRG-QSO connection, galaxies undergo an obscuration phase after the merger occurred due to an increase in star formation and dust production. This obscured phase is thought to be succeeded by a consecutive AGN phase during which feedback from the central SMBH repels the dust \citep{Hopkins2008}. 
As an illustration, the NGC 6240, low-ionization nuclear emission-line regions (LINER), is a well-studied \ul\ with high infrared luminosity, indicating a starburst activity that generates a large amount of dust and molecular gas \citep{Iono2007}. A detailed X-ray spectrum from BeppoSAX of NGC 6240 reveals a very strong absorption with $N_{H}$ $\sim$ 2 $\times$ 10$^{24}$ cm$^{-2}$ \citep{Vignati1999}. 
Deep Chandra observations provide high spatial resolution X-ray map of NGC 6240 and reveal the presence of a double AGNs system hidden in the core of the galaxy \citep{Komossa2003}. These observations are consistent with the idea that galaxies were formed as a consequence of a merger.

\begin{figure}
      \includegraphics[width=84mm]{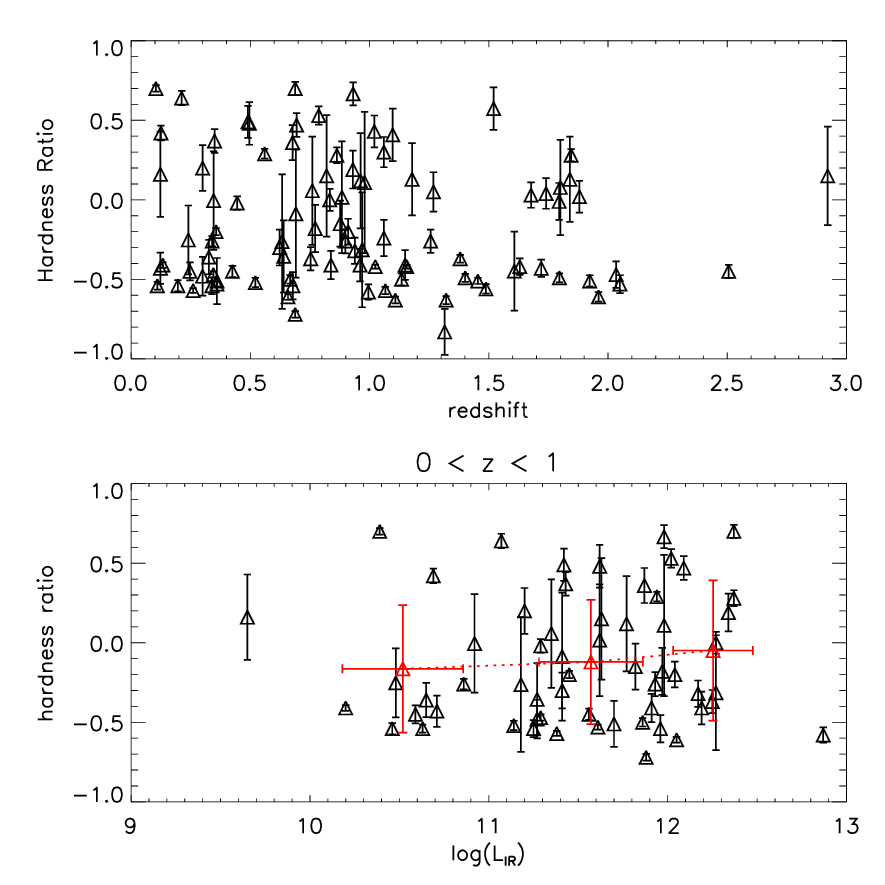}      
       \caption{Top Panel: hardness ratio as a function of redshift, \smx galaxies are labeled as black triangles. Bottom panel: hardness ratio versus log($L_{\rm IR}$) of \smx galaxies at the redshift range of 0$<$z$<$1. Red triangles represent the mean hardness ratio of \smx galaxies in three different log($L_{\rm IR}$) intervals. The x-axis and y-axis error bars are the 1-$\sigma$ standard deviation of hardness ratio and log($L_{\rm IR}$), respectively.}
  \label{fig:hr_z}
\end{figure}

To efficiently investigate the concomitant evolutionary stages of the AGN and star formation, our approach is to study the properties of galaxies detected in both infrared and X-ray.
\citet{Rafferty2011} had taken a similar approach to ours, looking for the X-ray counterpart of $70\micron$-selected galaxies in the ECDF-S, CDF-S, and EGS fields. 
They performed both a X-ray spectrum and X-ray band ratio analysis, assuming a power-law spectrum, and derived neutral hydrogen column densities for their 70$\micron$-selected X-ray and full X-ray selected AGN samples.
The $N_{H}$ from \citet{Rafferty2011} results ranged between 10$^{20}$ and 10$^{23}$ cm$^{-2}$, which is consistent with our mean value of $N_{H}$ $\sim$ 10$^{22}$ cm$^{-2}$. Both \citet{Rafferty2011} and our work have identified one possible Compton-thick AGN comparable to the X-ray properties of ULIRGs in the local universe. 
However, they did not identify any excess of neutral hydrogen in 70$\micron$-selected X-ray sources compared to other X-ray selected AGNs (Figure 4 in \citet{Rafferty2011} paper), in contrast with our result. 
Their result implies that the obscured AGN population is not higher among the 70$\micron$-selected X-ray sources, which contradicts the predictions of current AGN/starburst co-evolution models \citep{Hopkins2008}.
The main reason why \citet{Rafferty2011} have a different result to us, is probably that there are more ($\sim$ 15\%) \smx galaxies at 0$<$z$<$1 in our sample. 
In this redshift range, we find a large obscured AGN population (i.e. more AGNs with higher HR values; see the top panel of Figure~\ref{fig:hr_z}), which leads to the conclusion that \smx galaxies more often contain obscured AGNs. On the other hand, we lack $\sim$15\% \smx galaxies at the redshift z$>$2.
In Figure~\ref{fig:hr_z}, the top panel shows the HR as a function of redshift, higher values of HR are concentrated in the lower redshift bins. 
In contrast, the HRs of the higher redshift galaxies are relatively low. This result is expected because the rest-frame bands shift to higher frequency at high redshift, and so the HR becomes less sensitive to absorption.
The bottom panel of Figure~\ref{fig:hr_z} shows the HR versus log($L_{\rm IR}$) at 0$<$z$<$1. 
Given the large uncertainties, there is no apparent dependence between HR and log($L_{\rm IR}$). The mean HR is constant across the full log($L_{\rm IR}$) range.

\citet{Trichas2009} attempted a similar approach to our by studying the X-ray sources with $70\micron$ counterparts in the redshift range 0.5 $<$ z $<$ 1.3 from SWIRE survey. However, the K-S test resulted in a 3\% probability of X-ray sources with 70$\micron$ detection to be drawn from a purely X-ray selected sample. In fact their low statistic probably did not enable them to draw any robust conclusion on whether 70\micron\ detected X-ray galaxies were different from the global X-ray population. In contrast, we use the opposite selection method by selecting 70$\micron$ sources with X-ray counterparts and push to higher redshift (z $\sim$ 3) taking advantage of the COSMOS dataset. 
Owing to the twice times deeper 70$\micron$ observations in the COSMOS field we could indentify a larger \smx galaxy sample (142 galaxies) which enables us to perform a proper statistical study. We, therefore, take a different approach by applying a K-S test on the distribution of hardness ratio to investigate if \smx galaxies and X-ray selected AGNs are drawn from the same sample (see section~\ref{sec:hr-smx}).
Our K-S test concludes in less than 0.3\% of probability for \smx galaxies to be drawn from a purely X-ray selected sample.

\section{CONCLUSION}

We have investigated the properties of 142 galaxies both detected in X-ray and 70\micron\ in the COSMOS field. X-ray data are obtained from both XMM and Chandra point source catalogs, and 70\micron\ photometry is drawn from Spitzer-MIPS 70\micron\ point source catalog.  
We classify our sample into three distinct subsamples according to their respective total infrared luminosity ($L_{\rm IR}$): \sfgs\ (\lirsf), luminous infrared galaxies (\li s, \lirli), and ultra-luminous infrared galaxies (\ul s, \lirul), with median redshifts of z$\sim 0.168$, $0.518$ and $1.268$, respectively.
The major conclusions of this study are as follows:
\begin{enumerate}
\vspace{-0.2cm}
  \item We apply two methods to determine the physical mechanism shaping the SED of our objects, star formation or AGN: \\
  {\it a)} Using Spitzer-IRAC colours, we have shown that the majority (63\%) of our sample are classified as AGN. The AGN fraction is seen to increase with increasing total infrared luminosity ($L_{\rm IR}$).
  {\it b)} Using the relation between absorption corrected rest-frame hard X-ray luminosity $L_{\rm X}$ (2-10keV) and total infrared luminosity $L_{\rm IR}$ (8-1000$\micron$), we find that \smx galaxies have higher AGN activity relative to star formation. 
  
  \item Investigating the X-ray HR of our sample, we provide evidences for additional  X-ray absorption in the \smx galaxies compared to typical X-ray selected AGN samples, in agreement with the predictions of current AGN/starburst co-evolution models \citep{Hopkins2008}. The K-S tests on the HR distributions show that the probability for \smx galaxies and X-ray selected sample to be drawn from the same population is less than 0.3\%. The presence of more X-ray absorption suggests an additional process not accounted for in the conventional AGN unified model scheme.
  
  \item We find that there is no difference between \smx galaxies and 70$\micron$ galaxies in the observed far-infrared colour log($F_{70\micron}/F_{160\micron}$), suggesting the mechanism of far-infrared emission in galaxies could be mainly due to star formation disregarding the presence of AGNs.

\end{enumerate}

\section*{Acknowledgements}
We would like to appreciate the anonymous referee for his/her patience and constructive suggestions and corrections that further improved this paper. We also thank to Richard Davies for the corrections of the English text in my manuscript.
The work presented here is supported by the National Science Council of Taiwan under the grants NSC99-2112-M-003-002-MY2, NSC99-2119-M-003-005, and NSC99-2112-M-003-001-MY2.
We would like to acknowledge the wonderful work done by Jeyhan S. Kartaltepe on the Spitzer-COSMOS 70\micron\ point source catalog on which our analyses are based. We would like to thank the following people for very useful discussions, suggestions, and comments: Lin-Wen Chen, Jeyhan S. Kartaltepe, Wei-Hao Wang, Albert Kong, Matthew A. Malkan, Nick Scoville, David B. Sanders, Richard Davies, David Rosario, and Dieter Lutz. 
This research used extensively the COSMOS survey archive data of the NASA/IPAC Infrared Science Archive, which is operated by the Jet Propulsion Laboratory, California Institute of Technology, under contract with the National Aeronautics and Space Administration.

\end{document}